\documentclass[apj]{emulateapj}
\usepackage{apjfonts}
\usepackage{url}
\newcommand\nar{{New A Rev.}}

\begin{document}
\title{Extragalactic Background Light from Hierarchical Galaxy Formation: Gamma-ray Attenuation up to the Epoch of Cosmic Reionization and the First Stars}

\author{Yoshiyuki Inoue\altaffilmark{1}, Susumu Inoue\altaffilmark{2,3}, Masakazu A. R. Kobayashi\altaffilmark{4}, Ryu Makiya\altaffilmark{5},  Yuu Niino\altaffilmark{6}, \& Tomonori Totani\altaffilmark{5}} 

\affil{$^1$Kavli Institute for Particle Astrophysics and Cosmology, Department of Physics, Stanford University and SLAC National Accelerator Laboratory, 2575 Sand Hill Road, Menlo Park, CA 94025, USA}
\affil{$^2$Max-Planck-Institut f\"ur Kernphysik, Saupfercheckweg 1, 69117 Heidelberg, Germany}
\affil{$^3$Institute for Cosmic Ray Research, University of Tokyo, Kashiwa, Chiba 277-8582, Japan}
\affil{$^4$ Research Center for Space and Cosmic Evolution, Ehime University, Bunkyo-cho, Matsuyama 790-8577, Japan}
\affil{$^5$Department of Astronomy, Kyoto University, Sakyo-ku, Kyoto 606-8502, Japan}
\affil{$^6$ Optical and Infrared Astronomy Division, National Astronomical Observatory of Japan, Mitaka, Tokyo 181-8588, Japan}
\email{E-mail: yinoue@slac.stanford.edu}

\begin{abstract}
We present a new model of the extragalactic background light (EBL) and corresponding $\gamma\gamma$ opacity for intergalactic gamma-ray absorption from $z=0$ up to $z=10$, based on a semi-analytical model of hierarchical galaxy formation that reproduces key observed properties of galaxies at various redshifts. Including the potential contribution from Population III stars and following the cosmic reionization history in a simplified way, the model is also broadly consistent with available data concerning reionization, particularly the Thomson scattering optical depth constraints from {\it WMAP}. In comparison with previous EBL studies up to $z \sim3$--5, our predicted $\gamma\gamma$ opacity is in general agreement for observed gamma-ray energy below $400/(1 + z)$ GeV, whereas it is a factor of $\sim2$ lower above this energy because of a correspondingly lower cosmic star formation rate, even though the observed UV luminosity is well reproduced by virtue of our improved treatment of dust obscuration and direct estimation of star formation rate. The horizon energy at which the gamma-ray opacity is unity does not evolve strongly beyond $z\sim4$ and approaches $\sim20$ GeV. The contribution of Population III stars is a minor fraction of the EBL at $z=0$, and is also difficult to distinguish through gamma-ray absorption in high-$z$ objects, even at the highest levels allowed by the WMAP constraints. Nevertheless, the attenuation due to Population II stars should be observable in high-$z$ gamma-ray sources by telescopes such as {\it Fermi} or CTA and provide a valuable probe of the evolving EBL in the rest-frame UV. The detailed results of our model are publicly available in numerical form at the URL \url{http://www.slac.stanford.edu/\%7eyinoue/Download.html}.
\end{abstract}


\section{Introduction}
\label{sec:intro}
The extragalactic background light (EBL), the diffuse, isotropic background radiation from far-infrared (FIR) to ultraviolet (UV) wavelengths, is believed to be predominantly composed of the light from stars and dust integrated over the entire history of the Universe \citep[see][for reviews]{dwe12}. The observed spectrum of the local EBL at $z=0$ has two peaks of comparable energy density. The first peak in the optical to the near-infrared (NIR) is attributed to direct starlight, while the second peak in the FIR is attributed to emission from dust that absorbs and reprocesses the starlight.

The precise determination of the EBL has been a difficult task. Direct measurements of the EBL in the optical and NIR bands have been hampered by bright foreground emission caused by interplanetary dust, the so-called zodiacal light \citep[see][for reviews]{hau01}. Recently, \citet{mat11_pioneer} reported measurements of the EBL at 0.44 $\mu$m and 0.65 $\mu$m from outside the zodiacal region using observational data from {\it Pioneer 10/11}. On the other hand, integration over galaxy number counts provide a firm lower bound on the EBL, and the observed trend of the counts with magnitude indicates that the EBL at $z=0$ has been largely resolved into discrete sources in the optical/NIR bands \citep[e.g.][]{mad00,tot01,kee10}, even when the effect of incomplete detection due to cosmological dimming of surface brightness is taken into account \citep{tot01}.

The EBL can also be probed indirectly through observations of high-energy gamma rays from extragalactic objects \citep[e.g.][]{gou66,jel66,ste92,maz07}. Gamma-rays propagating through intergalactic space can be attenuated by photon-photon pair production interactions ($\gamma\gamma \rightarrow e^+e^-$) with low-energy photons of the EBL. For gamma-rays of given energy $E_\gamma$, the pair production cross section peaks for low-energy photons with energy
\begin{equation}
\label{eq:ene_ebl}
\epsilon_{\rm peak}\simeq \frac{2m_e^2c^4}{E_\gamma}\simeq0.5\left(\frac{1{\rm \ TeV}}{E_\gamma}\right)\ {\rm eV},
\end{equation}
where $m_e$ is the electron mass and $c$ is the speed of light. In terms of wavelength, $\lambda_{\rm peak} \simeq2.5(E_\gamma[{\rm TeV}])\ \mu{\rm m}$. Measuring the resultant attenuation features in the spectra of extragalactic GeV-TeV sources would offer a valuable probe of the EBL that is indirect, yet unique in being redshift-dependent. Although this method can be limited by incomplete knowledge of the intrinsic spectra of the source before attenuation, by assuming a plausible range for such spectra, observations of blazars by current ground-based telescopes have been able to place relatively robust upper limits to the EBL at $z=0$ and up to $z\sim0.5$ \citep[e.g.][]{aha06, alb08_3c279}. This has been complemented by {\it Fermi} observations of blazars and gamma-ray bursts (GRBs) that placed upper limits on the $\gamma\gamma$ opacity up to $z=4.35$ \citep{abd09_080916C,abd10_ebl}. The energy density of the local EBL has been constrained to be $<$ 24 nW m$^{-2}$ sr$^{-1}$ at optical wavelengths, and $<$ 5  nW m$^{-2}$ sr$^{-1}$ between 8 $\mu$m and 31 $\mu$m \citep{mey12}. Combined with the lower limits from galaxy counts, the total EBL intensity at $z=0$ from 0.1 $\mu$m to 1000 $\mu$m is inferred to lie in the range 52--99 nW m$^{-2}$ sr$^{-1}$ \citep{hor09}. Very recently, HESS has succeeded in positively measuring the imprint of the local EBL in the spectra of bright blazars, assuming only that their intrinsic spectra have smooth shapes \citep{abr13}. {\it Fermi} has also positively detected the redshift-dependent signature of EBL attenuation up to $z=1.5$, utilizing the collective spectra of a large number of blazars \citep{abd12}. However, the EBL at higher redshifts is still highly uncertain.

Currently available theoretical models for the EBL can be broadly categorized into three types. First, in backward evolution models, one starts from the observed properties of galaxies in the local Universe and describes their evolution by extrapolating backwards in time in a parameterized fashion \citep[e.g.][]{mal98,tot02,ste06,fra08}. This extrapolation entails uncertainties in the properties of the EBL that inevitably increase at high redshifts. Nevertheless, based on the observed, rest-frame K-band luminosity function (LF) of galaxies from $z=0$ up to $z=4$, \citet{dom11} were able to model the EBL without any assumptions for the LF. \citet{hel12,ste12} constructed evolving EBL models in a relatively robust way by utilizing multiwavelength photometric survey data.

Secondly, in forward evolution models, the basis is a description for the cosmic star formation history (CSFH), over which models for the spectral energy distribution (SED) of the stellar population are convolved to obtain the evolving EBL \citep[e.g.][]{kne04,fin10}.  However, such models cannot follow the detailed evolution of key physical quantities such as the metallicity and dust content, which can significantly affect the spectral shape of the EBL. Furthermore, although most forward evolution models employ the CSFH of \citet{hop06}, it is known that this CSFH model overproduces the stellar mass density \citep{far07,cho12}, and is also inconsistent with the observed rate of core-collapse supernovae \citep{hor11}. Recent studies by \citet{kob12} show that \citet{hop04,hop06} may have overestimated the CSFH at $z>1$, arising from overcorrection for dust obscuration effects and in conversion from luminosity to star formation rate.

Finally, rooted in the modern cosmological framework of large-scale structure formation driven by cold dark matter, semi-analytical models of hierarchical galaxy formation account for the merging history of dark matter halos as well as the physical evolution of the baryonic component, including the effects of gas cooling, star formation, metal enrichment, feedback heating, etc. \citep{pri05,gil09,you11,gil12}. Such models successfully reproduce various observed properties of galaxies from the local Universe up to $z\sim6$ \citep[see e.g.,][]{kau93,col94,nag99,som99,nag04,bau05,nag05,kob07,kob10,som12}. At present, semi-analytical models can be considered the most detailed and well-developed models for the EBL over a wide range of redshifts.

A subject that has yet to be fully explored in the context of the EBL and gamma-ray absorption is the epoch of cosmic reionization above $z\sim6$. Measurements of NIR absorption troughs in the spectra of high-$z$ quasars, together with those of anisotropies in the polarization of the cosmic microwave background (CMB), prove that the majority of intergalactic hydrogen in the Universe has been reionized somewhere between $z\sim30$ and $z\sim6$ \citep[see e.g.][]{rob10rev}. Although the most widely suspected source of reionization is UV photons emitted by early generations of massive stars, the observational constraints are still very limited, so that the actual sources, history and topology of cosmic reionization remain largely unknown. A closely related topic is the possible existence and formation history of Population III (Pop-III) stars, very massive stars that are expected to originate in nearly metal-free conditions, particularly for the very first generation of stars appearing in the Universe, and their potential role in cosmic reionization \citep[see e.g.][and references therein]{bro11,glo12}.

Since the current observational constraints on reionization  mostly concern the neutral or ionized intergalactic gas, it would be very valuable and complementary to obtain independent information on the evolving, UV intergalactic radiation field itself. A unique and promising approach may be offered by the effects of gamma-ray absorption in-situ in high-energy sources at $z>6$. UV radiation fields with sufficient intensities to cause cosmic reionization may induce significant gamma-ray absorption at observer energies above a few tens of GeV \citep{oh01,sin10}. Based on a semi-analytical model of galaxy formation that includes Pop-III stars and reproduces a variety of reionization-related observations\citep{cho06,cho09}, the recent study by \citet{sin10} suggested that appreciable attenuation may be expected above $\sim12$ GeV at $z\sim5$ and down to $\sim6-8$ GeV at $z\gtrsim8-10$, mainly caused by Pop-II stars at these epochs. However, the relative contribution of Pop-III stars was found to be difficult to discern observationally. On the other hand, without addressing the implications for reionization, some studies have concentrated on the prospects for constraining Pop-III star formation through gamma-ray absorption in objects at lower redshifts \cite[e.g.][]{gil11}.

The Fermi gamma-ray space telescope \citep[][{\it Fermi}]{atw09} may eventually detect blazars at $z>6$ \citep{ino10_highz},
and the Cherenkov Telescope Array \citep[][CTA]{act11} may possibly do the same for gamma-ray bursts (GRBs) \citep{sin12_cta}.
Therefore a deeper investigation into the above issues is worthwhile and timely. The above studies \citep{oh01,sin10,gil11} have not accounted consistently for the EBL resulting from galaxy formation at lower redshifts. For example, the model of \citet{sin10} was optimized to describe the reionization epoch and did not include the contribution from Pop-I stars or dust, and thus could only evaluate the gamma-ray opacity above $z=4$. 

In this paper, we present a new study of the EBL and consequent gamma-ray opacity, covering the entire redshift range from $z=0$ up to $z=10$ within a consistent framework, accounting for the process of cosmic reionization, and including Pop-III stars in a simplified way. As a baseline model, we adopt the Mitaka model\footnote{Named after the city of Mitaka where the model was mainly developed at the National Astronomical Observatory of Japan.} of semi-analytical galaxy formation \citep{nag04}.
The model can reproduce various observed properties of galaxies such as their luminosity function (LF), luminosity density (LD), and stellar mass density \citep{nag04}, as well as the LFs of high-redshift Lyman-break galaxies (LBGs) and Lyman-$\alpha$ emitters (LAEs) up to $z\sim6$ \citep{kob07,kob10}. As regards Pop-III stars, in view of the presently large theoretical uncertainties on their formation efficiency, metal production, conditions for transition to Pop-II star formation, etc., we choose not to fully incorporate them into our semi-analytical scheme. Instead, their formation history is characterized in a simple, parameterized way, which we constrain by modeling the cosmic reionization process and comparing with available observations, particularly the Thomson scattering optical depth measured by the Wilkinson Microwave Anisotropy Probe ({\it WMAP}).

We introduce our semi-analytical model in \S\ref{sec:mitaka}. Cosmic reionization is modeled and compared with observations in \S\ref{sec:reion}. \S\ref{sec:ebl} presents the results of our EBL models. The consequent gamma-ray opacity and comparison with current gamma-ray observations are described in \S\ref{sec:tau}. We conclude in \S\ref{sec:con}. Throughout this paper, we adopt the standard cosmological parameters of $(h, \Omega_M, \Omega_\Lambda) = (0.7, 0.3, 0.7)$, and a Salpeter initial mass function \citep[IMF;][]{sal55} within a mass range of 0.1 -- 60 $M_{\odot}$.

\section{Semi-analytical Galaxy Formation Model}
\label{sec:mitaka}

In the framework of the Mitaka semi-analytical model of galaxy formation, we follow the merger history of dark matter halos and the evolution of baryonic components. The evolution of the baryons within halos is modeled with physically motivated, phenomenological prescriptions for radiative cooling, star formation, supernova feedback, chemical enrichment, and galaxy merging. We can compute a variety of physical and observational quantities for individual galaxies as well as the global average over the Universe at any redshift, such as the CSFH, and LFs and dust content of galaxies. A mock catalog of galaxies can be generated that can be compared with different observations. More details of the Mitaka model are described in \citet{nag04,kob07,kob10}. Several free parameters in the prescriptions for baryons are fixed so that they fit a number of observed properties of local galaxies, such as their B-band and K-band LFs, neutral gas fraction, and gas mass-to-luminosity ratio as a function of B-band luminosity \citep{nag04}. For simplicity and consistency, we keep these parameters unchanged throughout this paper.

\subsection{Cosmic Star Formation History}

\begin{figure}
\begin{center}
\centering
\plotone{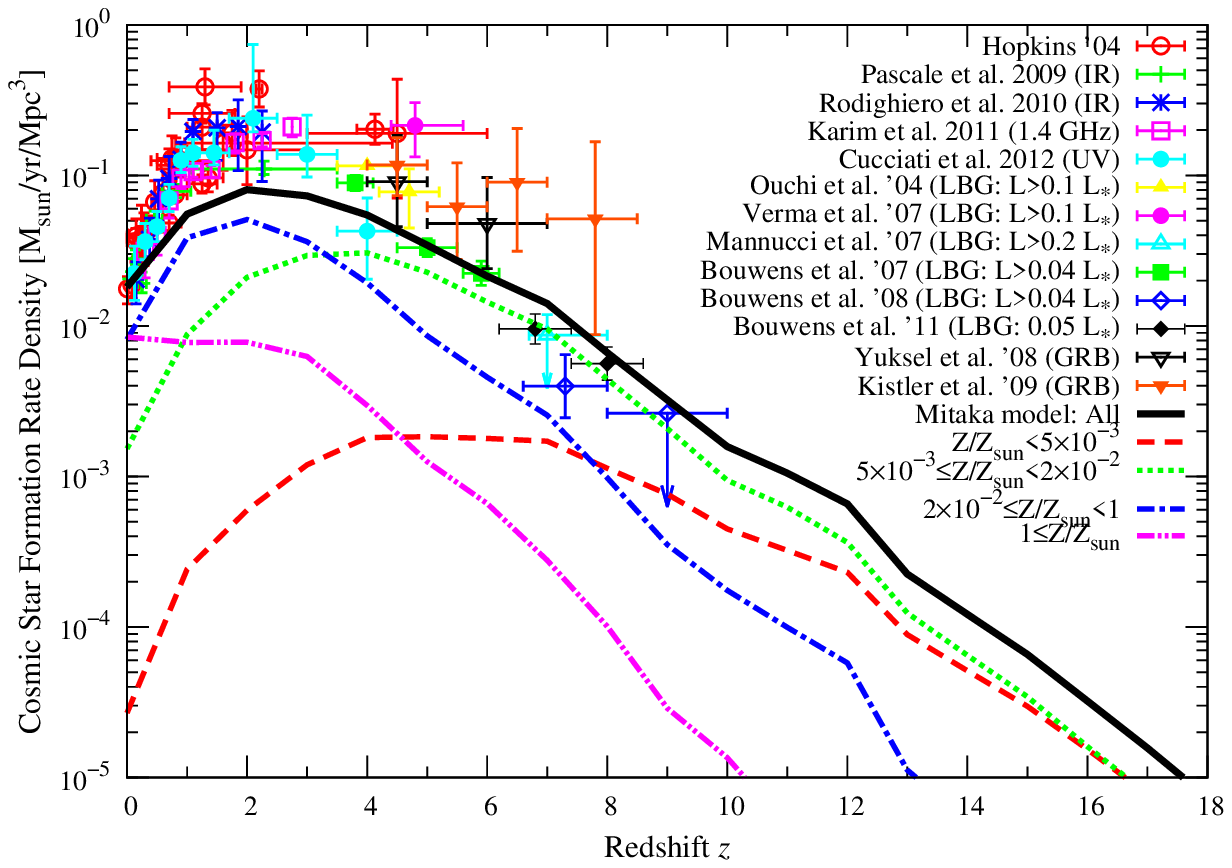} 
\caption{Cosmic star formation history.
Solid curve shows the total in the baseline Mitaka model, while the
dashed, dotted, dot-dashed and double-dot-dashed curves show the fractional contributions from stars with metallicity $Z/Z_{\odot}<5\times10^{-3}$, $5\times10^{-3} \le Z/Z_{\odot} < 2\times10^{-2}$, $ 2\times10^{-2} \le Z/Z_{\odot} < 1$, and $Z/Z_{\odot} \ge 1$, respectively. We also plot the observational data compiled by \citet{hop04}, that deduced from LBGs \citep{ouc04,bou07,man07,ver07,bou08}, and that inferred from GRBs \citep{yuk08,kis09}. For the LBG sample, limiting luminosities adopted by each author are indicated in the corresponding legend in term of the characteristic luminosity $L_*$ of the luminosity function. 
 \label{fig:csfh} }
\end{center}
\end{figure}

Fig. \ref{fig:csfh} shows the CSFH expected in the Mitaka model over $z=0-18$ in different ranges of metallicity. Pop-III stars are expected to form from gas with metallicity below a critical value, such that the gas can only cool rather inefficiently through rotational transitions of molecular hydrogen, which leads to fragmentation into relatively massive protostellar clouds, and ultimately the formation of very massive stars. Once the metallicity exceeds this value, the gas can cool more efficiently via metal emission lines, and a transition to the formation of less massive, Population II (Pop-II) stars is thought to take place \citep[e.g.][]{mac03,bro03,yos04}. However, the exact value of this critical metallicity has been debated, ranging from $Z=10^{-6}\  Z_\odot=10^{-7.7}$ \citep{sch06} to $Z=10^{-2}\  Z_\odot=10^{-3.7}$ \citep{ayk11}\footnote{We adopt $Z_\odot\simeq 0.02$ \citep{and89,gre98}, although an updated value of $Z_\odot \simeq 0.0134$ has been given by \citet{asp09}.}. In this paper, we consider stars with metallicity $Z<10^{-4}=5\times10^{-3}\ Z_\odot$ to correspond to Pop-III stars.  We adopt a Salpeter IMF in the mass range of $0.1-60M_{\odot}$ for all types of stars. Recent radiation-hydrodynamics simulations of Pop-III star formation suggest that their typical masses may be limited to $\lesssim40M_{\odot}$ due to radiative feedback effects \citep{hos11}, which would be in accord with our choice of the the maximum mass for Pop-III stars.

We also plot the data compiled by \citet{hop04, pas09, rod10, kar11,cuc12}, that deduced from LBGs, \citep{ouc04,bou07,man07,ver07,bou08,bou11_csfh}, and that inferred from GRBs \citep{yuk08,kis09}. Above $z>4$, each LBG data point is obtained by integrating its LF down to a certain limiting luminosity which is parameterized by the characteristic luminosity $L_*$ of the LF, whose choice often differ among authors. Our model shows the star formation in all galaxies, down to the faintest luminosities.

There is an apparent discrepancy between our model and the observed CSFH data at $1<z<5$. In our semi-analytical model, we directly estimate the star formation rate for each galaxy and evaluate the CSFH by integrating over all galaxies. The CSFH data points are converted from the observed galaxy LFs and involve uncertainties in the faint-end slope of the LF, dust obscuration correction from UV data, contamination from old stellar populations to the IR luminosity, total IR luminosity modeling, and a conversion factor from luminosity to star formation rate. For the CSFH parameterizations of
\citet{hop04} and \citet{hop06}, there are inconsistencies with other observational information such as the stellar mass density \citep[e.g.][]{cho12}, core collapse supernovae rate \citep{hor11} and constraints from gamma-ray observations \citep{rau12}. Recently, \citet{kob12} have shown that this discrepancy arises from overcorrection in dust obscuration and star formation rate conversion, which leads to a factor of $\sim$2--3 overestimation of the CSFH \citep{kob12}. The discrepancy does not adversely affect our results, since our model can reproduce various other observed properties of galaxies. The comparison of the CSFH correction methods between \citet{hop04} and \citet{kob12} is discussed in Appendix \ref{sec:app1}.

\subsection{Stellar and Dust Emission}

The cosmic emissivity due to stellar and reprocessed dust emission
at a given frequency $\nu$ and redshift $z$ is given by
the sum of their respective emissivities $j_{\rm star}(\nu,z)$ and $j_{\rm dust}(\nu, z)$,
\begin{equation}
j(\nu,z) = j_{\rm star}(\nu,z) +  j_{\rm dust}(\nu, z).
\end{equation}

We calculate $j_{\rm star}(\nu,z)$ from the CSFH using stellar population synthesis models that provide the SEDs as a function of metallicity and dust attenuation, namely the models of \citet{bru03} for $Z\ge10^{-4}$ and \citet{sch03} for $Z<10^{-4}$ (Pop-III);
note that the latter metallicity range is not covered by \cite{bru03}. We adopt the Salpeter IMF for the models with a correction for the IMF mass range used in our model. Thus,
\begin{eqnarray}
\nonumber
j_{\rm star}(\nu,z)&=&\int_z^{\infty} \left|\frac{dt}{dz'}\right|dz'\int_0^\infty dZ\int_0^{\infty} dA_\mathrm{V}f_{\rm esc}\dot\rho_{\rm star}(z',Z,A_V)\\
&\times& \varepsilon(\nu', z', z, Z)\exp[-\tau_{\rm ISM}(\nu',A_{\rm V})\times\tau_{\rm IGM}(\nu,z',z)],
\end{eqnarray}
where $A_\mathrm{V}$ is the interstellar dust attenuation strength in the {\it V}-band, $\dot\rho_{\rm star}(z,Z,A_V)$ is the CSFH for stars with metallicity $Z$ and dust attenuation $A_\mathrm{V}$ at redshift $z$ in units of ${\rm M}_{\odot}\ {\rm yr}^{-1}\ {\rm Mpc}^{-3}$, $\varepsilon(\nu,z',z,Z)$ is the intrinsic emissivity at frequency $\nu$ at $z$  from stars with metallicity $Z$ born at $z'$ in units of erg s$^{-1}$ Hz$^{-1}$ M$_{\odot}^{-1}$ given by stellar population synthesis models, and $\nu'=(1+z')\nu/(1+z)$. $\tau_{\rm ISM}$ and $\tau_{\rm IGM}$ are the attenuation opacities in the interstellar medium (ISM) and the intergalactic medium (IGM), respectively. We adopt the ISM dust attenuation law of \citet{cal00} and the IGM opacity of \citet{yos94}. $f_\mathrm{esc}$ is the escape fraction of photons from galaxies with energy above the threshold for ionization of hydrogen, $E=13.6$ eV; non-ionizing photons with lower energies are assumed to escape freely.

\begin{figure}
\begin{center}
\plotone{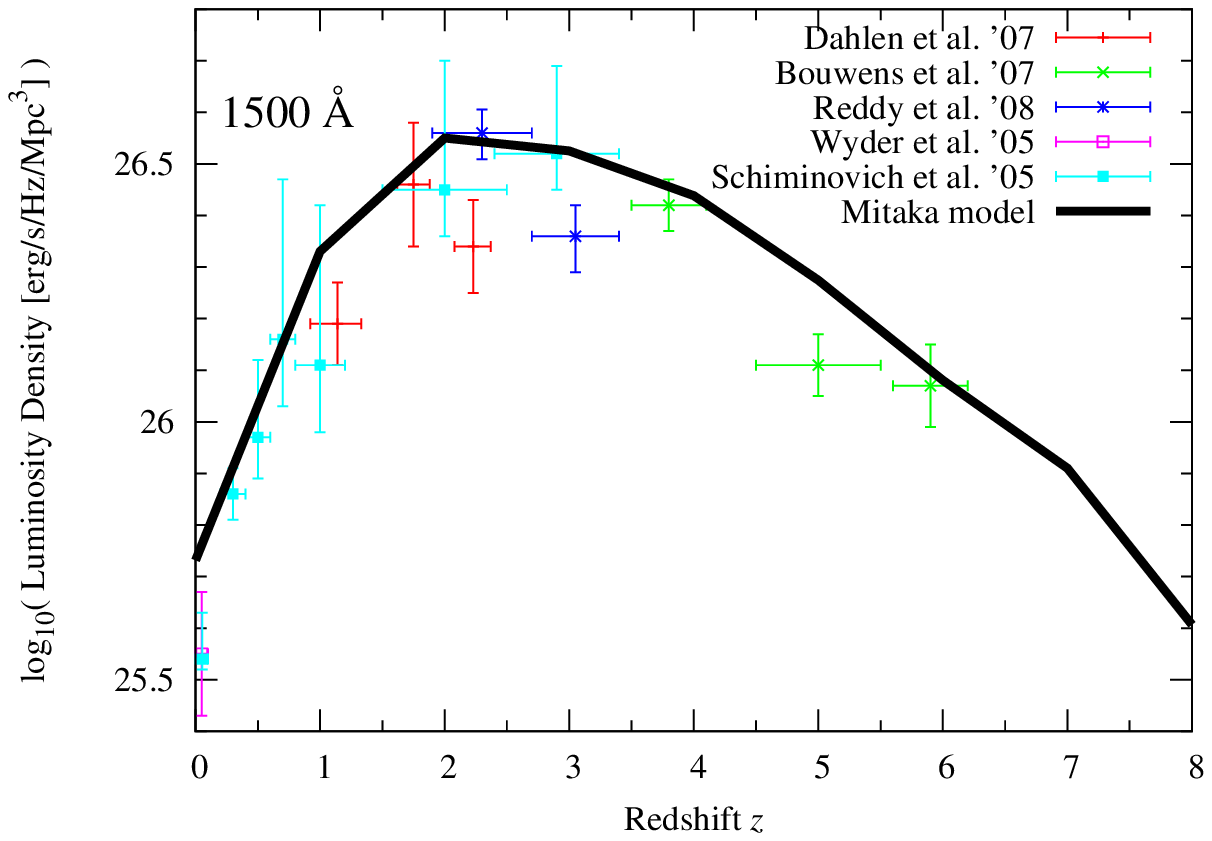} 
\caption{Luminosity density at rest-frame wavelength 1500 \AA. Solid curve is the result from the baseline Mitaka model. The observed data at various redshifts are also shown as indicated in the figure \citep{wyd05,sch05,dah07,bou07,red08}. \label{fig:LD} }
\end{center}
\end{figure}

Various observations have constrained the UV LD up to $z\sim6$  \citep{wyd05,sch05,dah07,bou07,red08}. Fig. \ref{fig:LD} compares the UV LD from our model with the observed LD data at a rest-frame wavelength of 1500 \AA. All the data agree well with our semi-analytical model. Comparison with other kinds of data are shown in \citet{nag04,nag05,kob07,kob10}.

Dust emits mid-IR (MIR; 5 $\sim30$ $\mu$m) and FIR ($\sim30\sim1000\ \mu$m) photons by reemitting the absorbed starlight. For dust emission, we also utilize a new implementation in the Mitaka model \citep[see][for details]{mak12}. We set the total dust emissivity $J_{\rm dust}(z)=\int_0^{\infty} d\nu j_{\rm dust}(\nu, z) $ to be equivalent to the total starlight energy absorbed by dust in the ISM,
\begin{eqnarray}
\nonumber
J_{\rm dust}(z)&=&\int d\nu\int_z^{\infty} \left|\frac{dt}{dz'}\right|dz'\int_0^\infty dZ\int_0^{\infty} dA_\mathrm{V}\dot\rho_{\rm star}(z',Z,A_V)\\
&\times& \varepsilon(\nu', z', z, Z)\{1-\exp[-\tau_{\rm ISM}(\nu',A_{\rm V})]\}.
\end{eqnarray}
To determine $j_{\rm dust}(\nu, z)$, we utilize the model of dust emission SED by \citet{dal02}, where the IR SED shape is defined by the exponent $\alpha$ of their Eq. 1. We adopt $\alpha=1.2$ for all galaxies in our model in order to reproduce the peak wavelength of the FIR EBL. Different $\alpha$ parameters result in different positions of the FIR peak wavelength.

Although our dust emissivity model can reproduce the local {\it Herschel} galaxy luminosity function \citep{vac10}, the redshift evolution of the IR luminosity function \citep[e.g.][]{rod10} had yet to be reproduced \citep{mak12}. Thus, we predict an IR EBL at $z=0$ that is lower than the current lower limits to the EBL by a factor of two. To remedy this, here we set the dust emissivity in our model to be three times more luminous than in the version of the Mitaka model by \citet{mak12}. Our model predictions in the MIR--FIR is therefore uncertain, while it should be more reliable in the UV-NIR band. We also note that the gamma-ray opacity due to the dust emission should be important only above several TeV in the local Universe (see Eq. \ref{eq:ene_ebl}).

\section{Cosmic Reionization History}
\label{sec:reion}

\begin{figure}
\begin{center}
\plotone{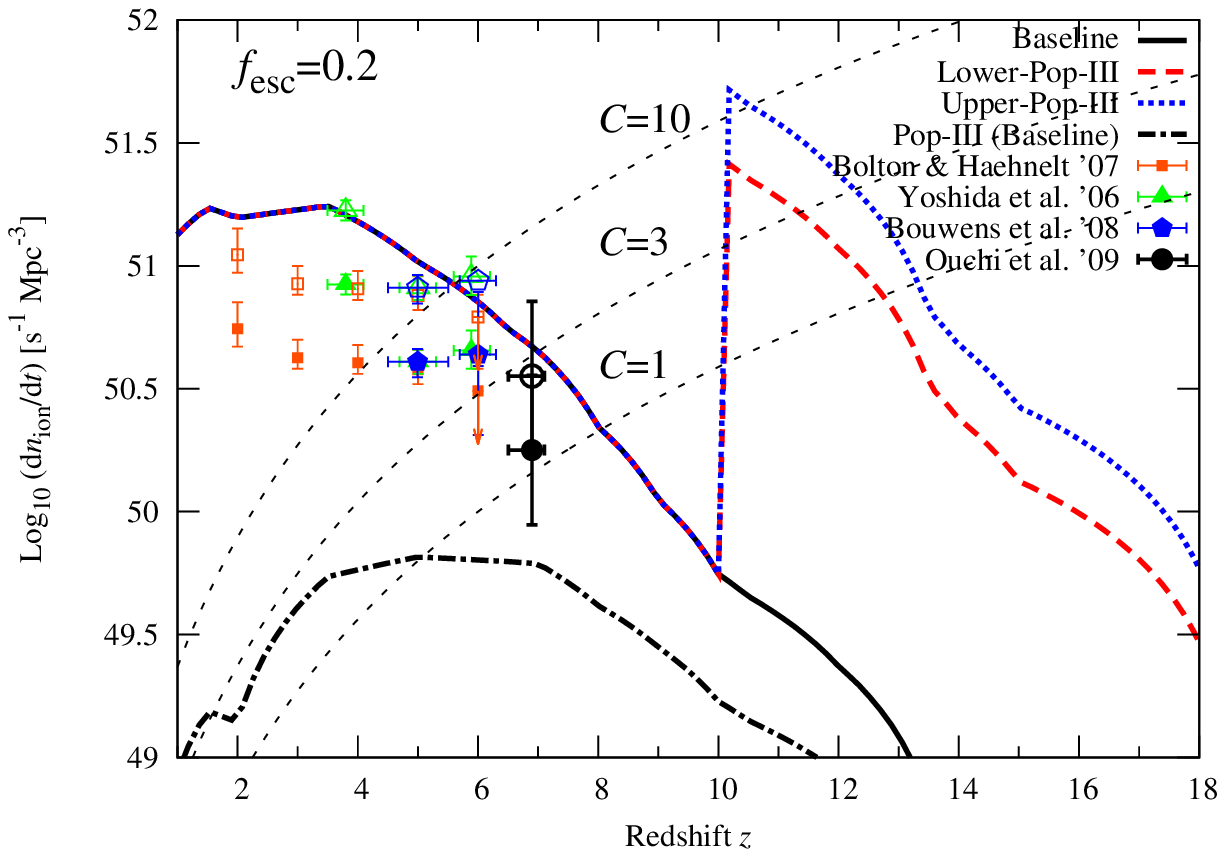} 
\caption{Ionizing photon emissivity per comoving Mpc$^3$, $dn_{\rm ion}/dt$, as a function of redshift. We set $f_{\rm esc}=0.2$ for all models and observed data. Solid, dashed, and dotted curves shows the baseline, $(\xi=1.0$ and $z_c=0.0$), the lower-Pop-III, , $(\xi=50.0$ and $z_c=10.0$), and the upper-Pop-III model $(\xi=100.0$ and $z_c=10.0$), respectively. Dot-dashed curve shows the ionizing photon emissivity from Pop-III population in the baseline model. The triangle, pentagon, and circle symbols show the data derived from the UV LFs of galaxies by \citet{yos06}, \citet{bou08}, and \citet{ouc09}, respectively. The data derived from the combination of hydrodynamical simulations and Ly$\alpha$ forest opacity \citep{bol07} are shown by square symbols. Filled and open symbols correspond to $\alpha_{\rm ion}=3$ and $1.5$, respectively, for the spectral index of ionizing emission. Thin dashed lines plot the estimated ionizing photon emissivity that is required for maintaining the ionization of hydrogen in the IGM \citep{mad99} for clumping factors of $C=1$, 3, and 10, from bottom to top. \label{fig:nion} }
\end{center}
\end{figure}

After the epoch of cosmic recombination, the Universe entered the so-called dark ages, a period with no significant sources of radiation. As the initially small fluctuations in the matter density field grew by gravitational instability and collapsed to form dark matter halos, the baryons that fell into sufficiently massive halos are expected to have cooled efficiently to form the first stars and galaxies \citep{bro11,glo12}. Such stars and galaxies should generate UV radiation that ionize their environments to create the first H II regions, which eventually grow and overlap to reionize the entire intergalactic medium. Observationally, cosmic reionization is known to have proceeded at least partially by $z \sim 10$ and been essentially completed by $z \sim 6$. However, the actual history, nature and sources of reionization are still largely unconstrained \citep[see][for reviews]{bar01,fan06}. In this section, we discuss how we model the reionization history of the Universe with our Mitaka model.

\subsection{Ionizing Photon Emission Rate}
First we evaluate the emissivity of photons with energies greater than 13.6 eV that can ionize hydrogen atoms. A key uncertainty is the escape fraction $f_{\rm esc}$ of ionizing photons from galaxies, which we assume here to be a constant value of 0.2 at all redshifts. This is motivated by the numerical simulations of \citet{yaj09,yaj11}, although they also showed that $f_{\rm esc}$ can depend on halo mass. Observationally, $f_{\rm esc}\simeq0.05$ is found in LBGs at $z\sim3$ \citep{sha06,iwa09}, but values at $z\ge4$ have not been determined yet. \citet{ono10} have set upper limits of $f_{\rm esc}\lesssim0.6$ at $z=5.7$ and $f_{\rm esc}\lesssim0.9$ at $z=6.6$ for LAEs.

Fig. \ref{fig:nion} shows the ionizing photon emissivity $dn_{\rm ion}/dt$ in units of s$^{-1}$ Mpc$^{-3}$, compared with various observations. We set $f_{\rm esc}=0.2$ for interpretation of all the observed data as well. The observed ionizing emissivities were derived by \citet{ouc09} from galaxy UV LFs at $z=4-7$ \citep[][filled symbols in Fig. \ref{fig:nion}]{yos06,bou08,ouc09}. The conversion from LF to ionizing photon emissivity is based on Eq. 5 in \citet{ouc09}, where continuous star formation history is assumed for all galaxies. The data are integrated down to $L=0$. We also show the ionizing photon rate inferred from the Ly$\alpha$ forest by combining hydrodynamical simulations with measurements of the Ly$\alpha$ opacity of the IGM \citep[][open symbols in Fig. \ref{fig:nion}]{bol07}.

As shown later, the baseline Mitaka model does not produce enough ionizing photons to account for the Thomson scattering optical depth measured by {\it WMAP}, although it can reionize the Universe sufficiently at $z\lesssim8$. In order to achieve consistency with observations, we extend the baseline model by considering an additional potential contribution of ionizing photons from Pop-III stars in a simplified way. Although ideally one would like to incorporate Pop-III stars self-consistently into our semi-analytic scheme, this is currently precluded by very large uncertainties in their formation efficiency, metal yield, etc. Instead we introduce two new parameters and simply enhance the total ionizing photon emissivity of the baseline Mitaka model by a constant factor $\xi$ above a critical redshift $z_c$, and attribute such an additional component to Pop-III stars. An alternative procedure might be to enhance the emissivity of only the stellar population with $Z<10^{-4}$ in the baseline model (Figs.\ref{fig:csfh} and \ref{fig:nion}), but this will not be more satisfactory in any way as the actual evolution of the metallicity would be altered. Thus we choose to simply enhance the total ionizing emissivity. In Fig. \ref{fig:nion}, we show the cases with $(\xi,z_c)=(50.0, 10.0)$ and $(50.0, 10.0)$,
referred to as the lower-Pop-III model and the upper-Pop-III model, respectively. Also shown is the contribution of stars with $Z<10^{-4}$ in the baseline model with $(\xi,z_c)=(1.0,0.0)$. The corresponding star formation rates of Pop III stars at $z=10$ will be $4.5\times10^{-4}$, $4.4\times10^{-2}$, and $8.8\times10^{-2}$ $\mathrm{M_\odot\ Mpc^{-3} \ yr^{-1}}$ for the baseline, the lower-Pop-III, and the upper-Pop-III model, respectively.

Although the overall behavior of our ionizing photon emissivity at $2<z<7$ is similar to that derived from the observed data, we overpredict the ionizing photon emissivity by about a factor of 2. We note that the data points are very sensitive to the assumed spectral index of the ionizing emission $\alpha_{\rm ion}$ \citep[see Eq. 5 in][]{ouc09}, which is set to be 3.0 for the data in Fig. \ref{fig:nion} but is actually not well determined. If $\alpha_{\rm ion}=1.5$, it will double the data derived from the galaxy UV LF and the Ly$\alpha$ forest opacity, bringing it into closer agreement with the model.

Fig. \ref{fig:nion} also shows $dn_{\rm ion}/dt$ that is required to balance the recombination of intergalactic hydrogen based on the formulation of \citet{mad99},
\begin{equation}
\label{eq:madau}
\frac{dn_{\rm ion}}{dt} = \frac{n_{\rm H}^0}{t_{\rm rec}(z)} \simeq 10^{47.4} C (1+z)^3 \ {\rm [s^{-1}Mpc^{-3}]}
\end{equation}
where $n_{\rm H}^0$ is the total number density of intergalactic hydrogen atoms (in both HI and H{II} phases), $t_{\rm rec}(z)$ is the recombination time scale at $z$, and $C=\langle n_H^2\rangle/\bar n_H^2$ is a time-dependent,volume-averaged clumping factor, for which the cases of $C=1,3,$ and 10 are shown. Note that $C=1$ corresponds to a uniform IGM. All our models have a sufficient budget of photons to ionize the Universe at $z\le7-8$. However, the baseline model can not {do so above $z=8$, in contradiction with WMAP observations that constrain the reionization redshift to be $z=10.6\pm1.2$ if it was instantaneous \citep{kom11}. Even if reionization occurred gradually, more ionizing photons may actually be necessary above $z\sim8$ than is implied by the estimates of \citet{mad99}.

\subsection{Probing the Cosmic Reionization History}

Important observational indicators of the reionization history are the optical depth to electron scattering and the neutral fraction of intergalactic hydrogen. Following \citet{bar01}, we compute the reionization history of the Universe. The equation of ionization equilibrium in terms of the volume filling factor $Q_{\rm HII}$ of HII regions is given by
\begin{equation}
\label{eq:reion}
\frac{dQ_{\rm HII}}{dt}=\frac{1}{n_H^0}\frac{dn_{\rm ion}}{dt} - \alpha_{B}\frac{C}{a(t)^3} n_H^0Q_{\rm HII},
\end{equation}
where $t$ is the cosmic time, $n_H^0=X_Hn_B^0$ is the present-day number density of hydrogen with $n_B^0$ as the present-day baryon number density and $X_H=0.76$ as the mass fraction of hydrogen, $dn_{\rm ion}/dt$ is the production rate of ionizing photons (see Fig. \ref{fig:nion}), and $\alpha_B=2.6\times10^{-13}\ {\rm cm^3s^{-1}}$ is the recombination rate of hydrogen at temperature $T=10^4$ K. The recombination time scale $t_{\rm rec}(z)$ in Eq. \ref{eq:madau} is given by $a(t)^3 /\alpha_{B}Cn_H^0$, where $a(t)$ is the cosmic scale factor.

Assuming a constant clumping factor $C$, Equation \ref{eq:reion} can be solved to give \citep[see][for details]{bar01}
\begin{equation}
Q_{{\rm HII}}(z_0) = \int_{z_0}^{\infty}dz\left|\frac{dt}{dz}\right|\frac{1}{n_H^0}\frac{dn_{\rm ion}}{dt}e^{F(z,z_0)},
\end{equation}
where $dt/dz$ is calculated from the Friedmann equation in the standard, flat universe cosmology as
\begin{equation}
\frac{dt}{dz}=\frac{1}{(1+z)H_0\sqrt{\Omega_M(1+z)^3+\Omega_\Lambda}}.
\end{equation}
Once $Q_{\rm HII}$ reaches 1, the IGM is fully ionized, and ionizing photons propagate freely in intergalactic space. The function $F(z,z_0)$ accounts for recombination and is given by
\begin{equation}
F(z,z_0)=-\frac{2}{3}\frac{\alpha_Bn_H^0}{\sqrt{\Omega_M}H_0}C[f(z)-f(z_0)],
\end{equation} 
where $f(z)$ is defined as
\begin{equation}
f(z)=\sqrt{(1+z)^3+\frac{1-\Omega_M}{\Omega_M}}.
\end{equation}
For the purpose of calculating $Q_{\rm HII}$, we set $C=3.0$, motivated by the numerical simulations of \citet{paw09}. We do not change $C$ as a function of redshift. We also assume $Q_{{\rm He II}}=Q_{{\rm HII}}$ for the volume filling factor of HeII regions and neglect the free electrons in HeIII regions for computing the optical depth to electron scattering. The number density of free electrons at $z$ is then
\begin{eqnarray}\nonumber
n_e(z)&=&\left(Q_{{\rm HII}}(z)X_H + \frac{Q_{{\rm He II}}(1-X_H)}{4}\right)n_B^0(1+z)^3\\
&=&\frac{1+3X_H}{4}Q_{{\rm HII}}(z)n_B^0(1+z)^3.
\end{eqnarray}

\begin{figure}
\begin{center}
\plotone{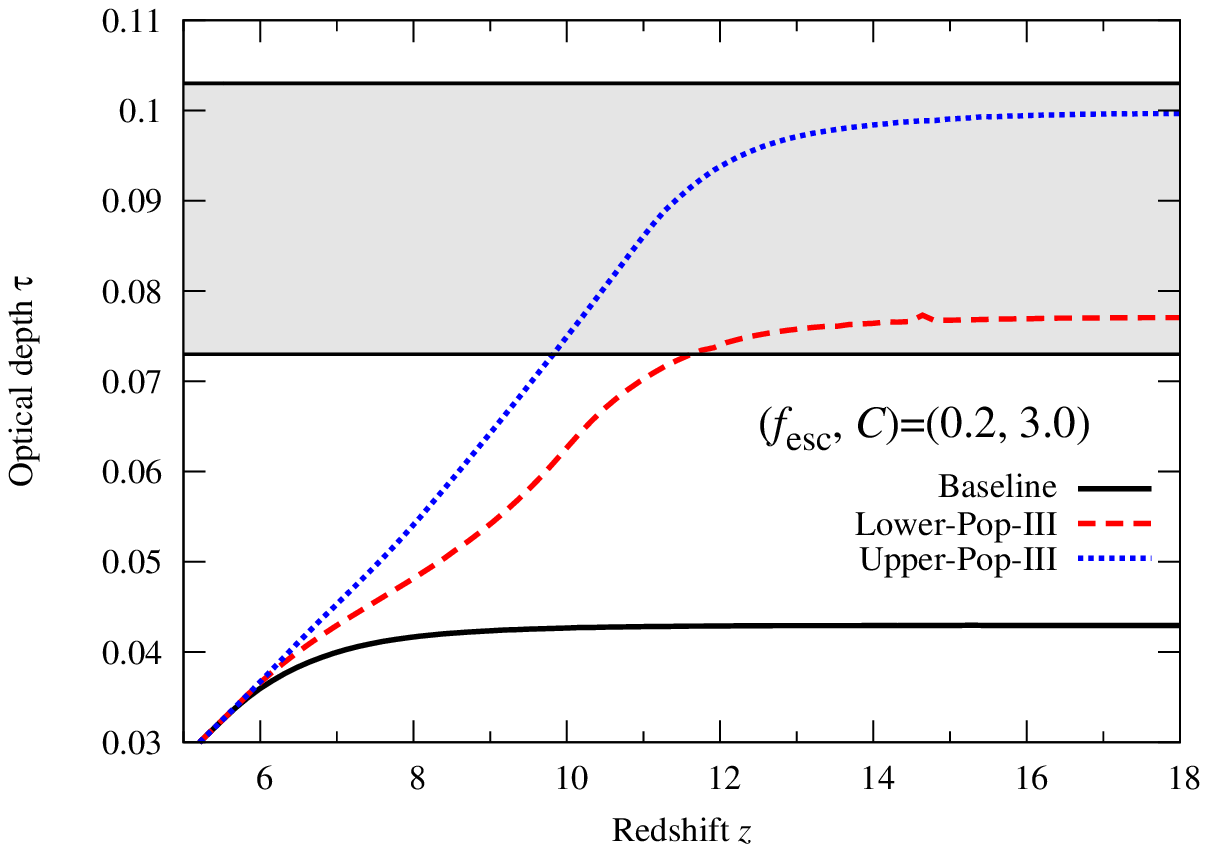} 
\caption{Thomson scattering optical depth of the IGM. We set $f_{\rm esc}=0.2$ and $C=3.0$. Solid, dashed, and dotted curves correspond to the baseline, lower-Pop-III, and upper-Pop-III models, respectively. The shaded region shows the 7-year {\it WMAP} results with 1-$\sigma$ errors, $\tau_e=0.088\pm0.015$ \citep{kom11}. \label{fig:tau_WMAP} }
\end{center}
\end{figure}

The optical depth to electron scattering is 
\begin{equation}
\tau_e(z_0)=\int_0^{z_0}dz\frac{dl}{dz}\sigma_Tn_e(z),
\end{equation}
where $\sigma_T$ is the Thomson cross section and $dl/dz$ is the cosmological line element for a standard, flat universe cosmology given by 
\begin{equation}
\frac{dl}{dz}=c\frac{dt}{dz}=\frac{c}{(1+z)H_0\sqrt{\Omega_M(1+z)^3+\Omega_\Lambda}}.
\end{equation}

Fig. \ref{fig:tau_WMAP} shows the Thomson scattering optical depth of the Universe, together with the range of $\tau_e=0.088\pm0.015$ derived from the 7-year {\it WMAP} data \citep{kom11}. As mentioned above, the baseline model can not reproduce the {\it WMAP} data, despite managing to reionize the Universe at $z\lesssim8$. The results of the lower-Pop-III and upper-Pop-III models are close to the lower and upper limits from {\it WMAP}, respectively, implying that 50-100 times more ionizing photons are necessary at $z \gtrsim 10$ than is conservatively expected from our semi-analytical galaxy formation model that successfully accounts for various observations at $z \la 8$. The fact that the ionizing photon budget estimated from galaxy populations directly observed so far are insufficient to account for the WMAP $\tau_e$
is well documented \citep[e.g.][]{sta07,cha08,oes09,ouc09,paw09,bun10,lab10,rob10,bou11,bou12}.

\begin{figure}
\begin{center}
\plotone{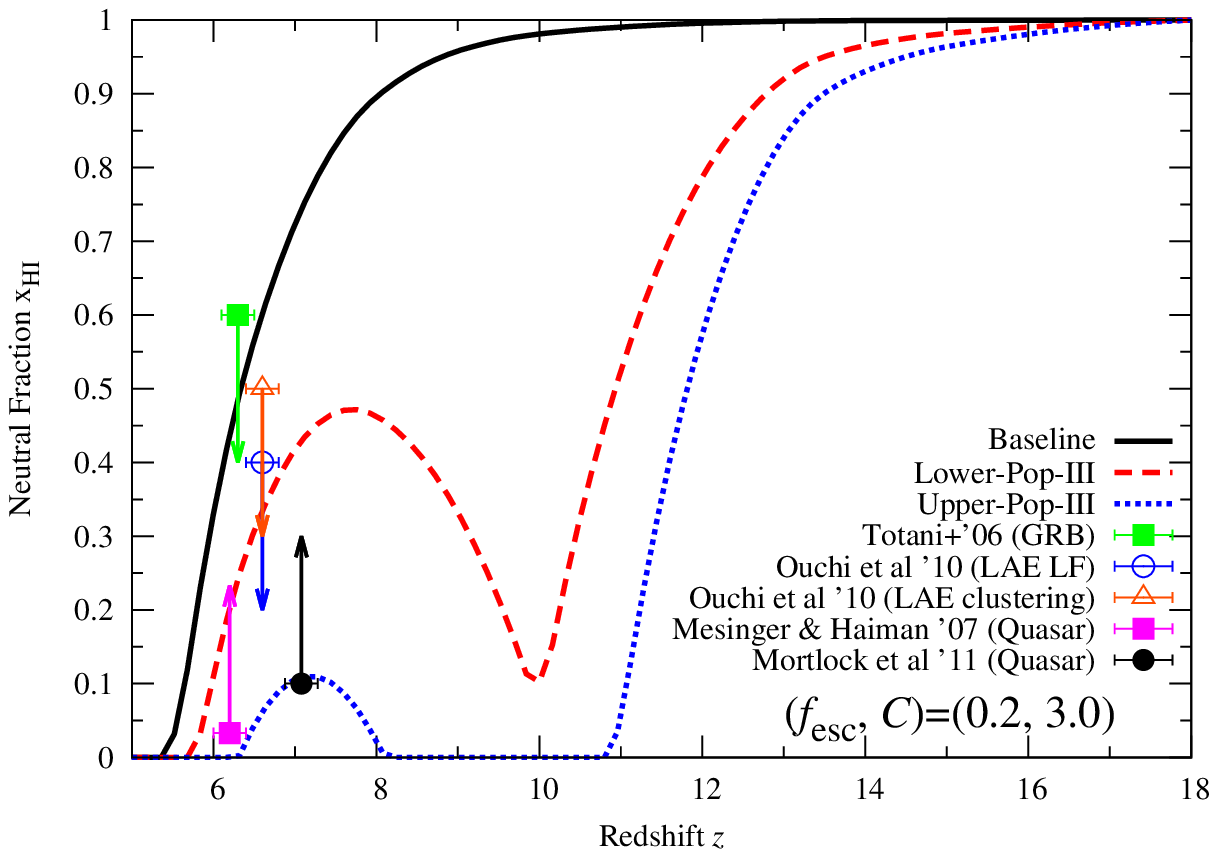} 
\caption{Neutral fraction of intergalactic hydrogen. We set $f_{\rm esc}=0.2$ and $C=3.0$. Solid, dashed, and dotted curves correspond to the baseline, lower-Pop-III, and upper-Pop-III models, respectively. We also show observational constraints from quasars \citep{mes07,mor11}, a GRB \citep{tot06}, LAE LFs  \citep{ouc10} and LAE clustering \citep{ouc10}. 
\label{fig:xH} }
\end{center}
\end{figure}

Fig. \ref{fig:xH} shows $x_{\rm HI}=1-Q_{\rm HII}$, the neutral fraction of intergalactic hydrogen, compared with constraints from analysis of Gunn-Peterson (GP) troughs \citep{gun65} in the spectra of quasars \citep{mes07,mor11} and a GRB \citep{tot06}. Also shown are constraints from LFs and clustering amplitudes of LAEs by \citep{ouc10}. They derived $x_{\rm H}<0.40$ at 1$\sigma$ confidence level by comparing the observed LAE LFs at $z=6.6$ with a theoretical model including Ly$\alpha$ transmission through the IGM,
but the limit varies from 0.1 to 0.53 depending on the model \citep[see \S 6.1.1. in][for details]{ouc10}. They also derived $x_{\rm H}\lesssim0.50$ by comparing the angular correlation functions and bias of their LAE samples at $z=6.6$ with the theoretical predictions of \citet{mcq07} and \citet{fur06}.

\begin{figure*}
\begin{center}
\plotone{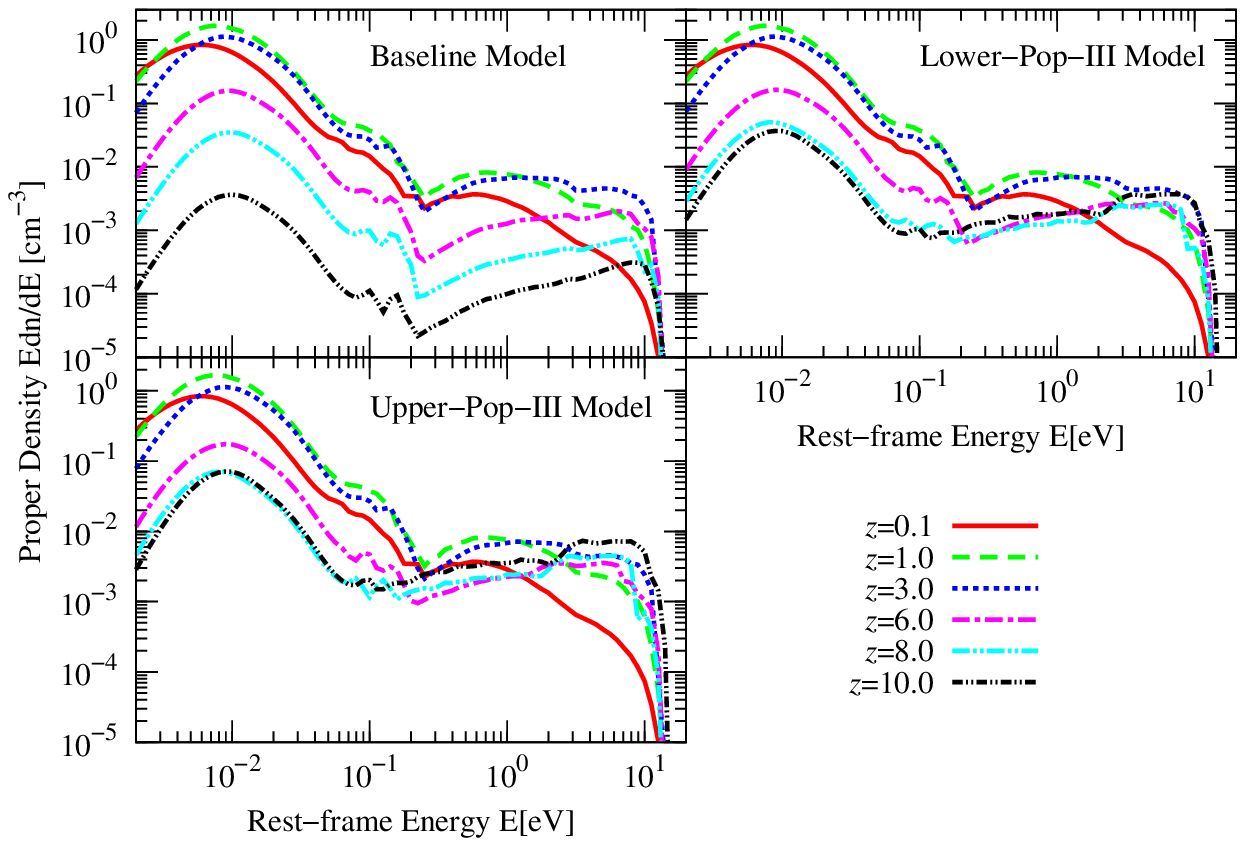} 
\caption{Proper volume photon number densities multiplied by the photon energy $\epsilon$ as a function of redshift. Top-left, top-right, and bottom-left panels correspond to the baseline, upper-Pop-III, and lower-Pop-III models, respectively. Solid, dashed, dotted, dot-dashed, double dot-dashed, and triple dot-dashed curves correspond to the proper photon density at $z=0.1$, 1.0, 3.0, 6.0, 8.0, and 10.0, respectively. \label{fig:ebl_prop} }
\end{center}
\end{figure*}

We have not attempted a detailed comparison with GP measurements of quasars at $z\lesssim6$ \citep[e.g.][]{fan06}. Such effects depend rather sensitively on the detailed distribution of regions with low neutral gas density, which is not essential for our purposes of modeling the EBL. 

Our baseline model that was inconsistent with {\it WMAP} is also seen to contradict the LAE constraints on the neutral fraction \citep{ouc10} (which are rather model-dependent as discussed above). In contrast, both the lower-Pop-III and the upper-Pop-III models are generally consistent with the current observational limits, although the latter may be in marginal conflict with the GP constraints of \citet{mes07}. Our simplifying assumption of a large enhancement of the ionizing photon emissivity only above $z = z_c$ in these two models leads to their nontrivial evolution of $x_{\rm HI}$, with a dip at $z \sim 10$ due to reionization by Pop III stars alone, followed by a peak at $z \sim 7-8$ due to partial recombination after Pop III termination, and then finally complete reionization by Pop II stars (c.f. \cite{cen03}). More realistic modeling with a smoother transition from Pop-III to Pop-II populations may make such features less pronounced.

Besides a large contribution from Pop-III stars,
we note that various other aspects may be important in achieving sufficient ionizing photons at $z \ga 8$ to account for the {\it WMAP} data
\citep[e.g.][]{haa12,kuh12}. These include steeping of the faint-end slope of the LF \citep{bou12}, smaller clumping factor \citep{bol07,paw09}, larger escape fraction \citep{yaj11}, harder initial mass function \citep{mck08},
and X rays from accreting black holes \citep{ric04,mir11}.

\section{Extragalactic Background Light}
\label{sec:ebl}
The background intensity $I(\nu_0,z_0)$ at redshift $z_0$ and frequency $\nu_0$ is computed by integrating the radiation from all sources between $z=z_0$ and the maximum redshift of the source distribution $z_{\rm max}$  \citep[see e.g.][]{pea99},

\begin{equation}
I({\nu_0},z_0)=\frac{1}{4\pi}\int_{z_0}^{z_{\rm max}}dz\frac{dl}{dz}j(\nu,z),
\label{eq:local_EBL}
\end{equation}
where $j(\nu,z)$ is the comoving volume emissivity at redshift $z$ and frequency $\nu=\nu_0(1+z)$, calculated by combining our CSFH and stellar population synthesis models. We set $z_{\rm max}=20$.

\begin{figure*}
\begin{center}
\epsscale{0.80}
\plotone{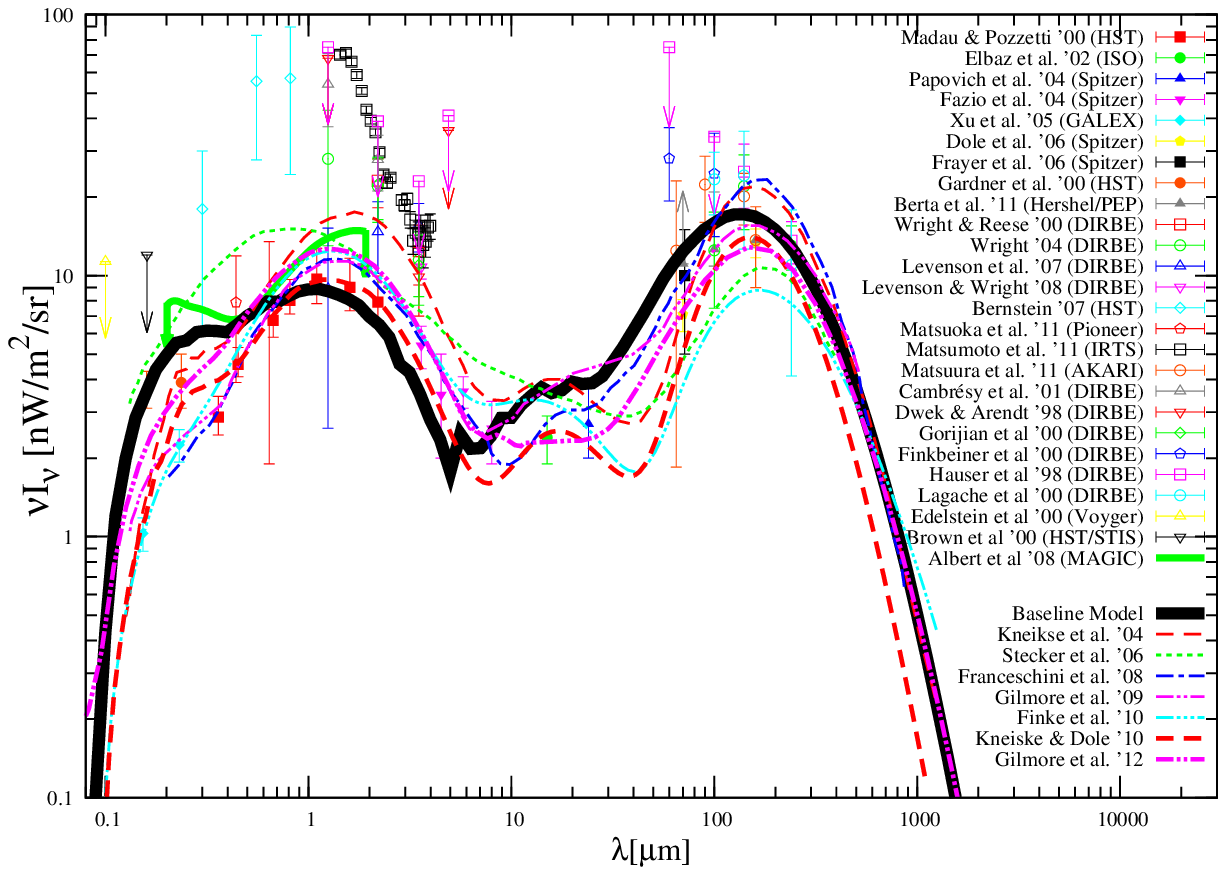} 
\caption{The EBL for the baseline model is shown by the solid curve. For comparison, the EBL models by \citet[thin dashed]{kne04}, \citet[dotted]{ste06}, \citet[dot-dashed]{fra08}, \citet[thin double dot-dashed]{gil09}, \citet[triple dot-dashed]{fin10}, \citet[thick dashed]{kne10}, and \citet[thick double dot-dashed]{gil12} are shown as indicated in the figure. The integrated brightness of galaxies (minimum EBL; filled symbols) and current measurements of the EBL (open symbols) are shown as indicated in the figure. References for the integrated brightness of galaxies are {\it HST} \citep{mad00,gar00}, {\it ISO} \citep{elb02}, {\it Spitzer} \citep{pap04,faz04,dol06,fra06}, {\it GALEX} \citep{xu05}, and {\it Hershel} \citep{ber11}. References for the current EBL measurements are {\it DIRBE} \citep{wri00,wri04,lev07,lev08,cam01,dwe98,gor00,fin00,hau98,lag00}, {\it HST} \citep{ber07,bro00}, {\it Pioneer} \citep{mat11_pioneer}, {\it IRTS} \citep{mat05}, {\it AKARI} \citep{mat11_akari}, and {\it Voyger} \citep{ede00}. The upper limits from TeV gamma-ray observations by MAGIC \citep{alb08_3c279} are also shown by the solid curve flanked by arrows. \label{fig:ebl} }
\end{center}
\end{figure*}

From Eq. \ref{eq:local_EBL}, the specific radiation energy density (in units of erg s$^{-1}$ cm$^{-3}$ Hz$^{-1}$) in the proper volume is 
\begin{equation}
\rho(\nu_0,z_0)=\frac{4\pi}{c}(1+z_0)^3I(\nu_0,z_0)
\end{equation}
The photon proper number density is 
\begin{equation}
\frac{dn(\epsilon_0,z_0)}{d\epsilon_0}=\frac{\rho(\nu_0,z_0)}{\epsilon_0},
\end{equation}
where $\epsilon_0=h_p\nu_0$ is the photon energy and $h_p$ is the Planck constant.

Fig. \ref{fig:ebl_prop} plots the proper photon number density for our models. The proper photon number density increases from $z=10$ up to $z\sim1-3$ where the CSFH reaches a peak and then decreases toward the local Universe. Once we increase the Pop-III component, the UV photon density at high redshifts becomes higher and comparable to that at $z=1$. Although the IR photon density also increases, it is not as significant as the UV since there is little dust at high redshifts.

The EBL intensity at $z=0$ for the baseline model is displayed in Fig. \ref{fig:ebl} as a function of wavelength $\lambda$. We also show other theoretical models for comparison \citep{kne04,ste06,fra08,gil09,fin10,kne10,gil12}, together with current measurements of the EBL and the integrated brightness of galaxies. Detailed predictions for the proper photon number density and the local EBL intensity is publicly available at our website\footnote{ \url{http://www.slac.stanford.edu/\%7eyinoue/Download.html}}.

\begin{figure}
\begin{center}
\plotone{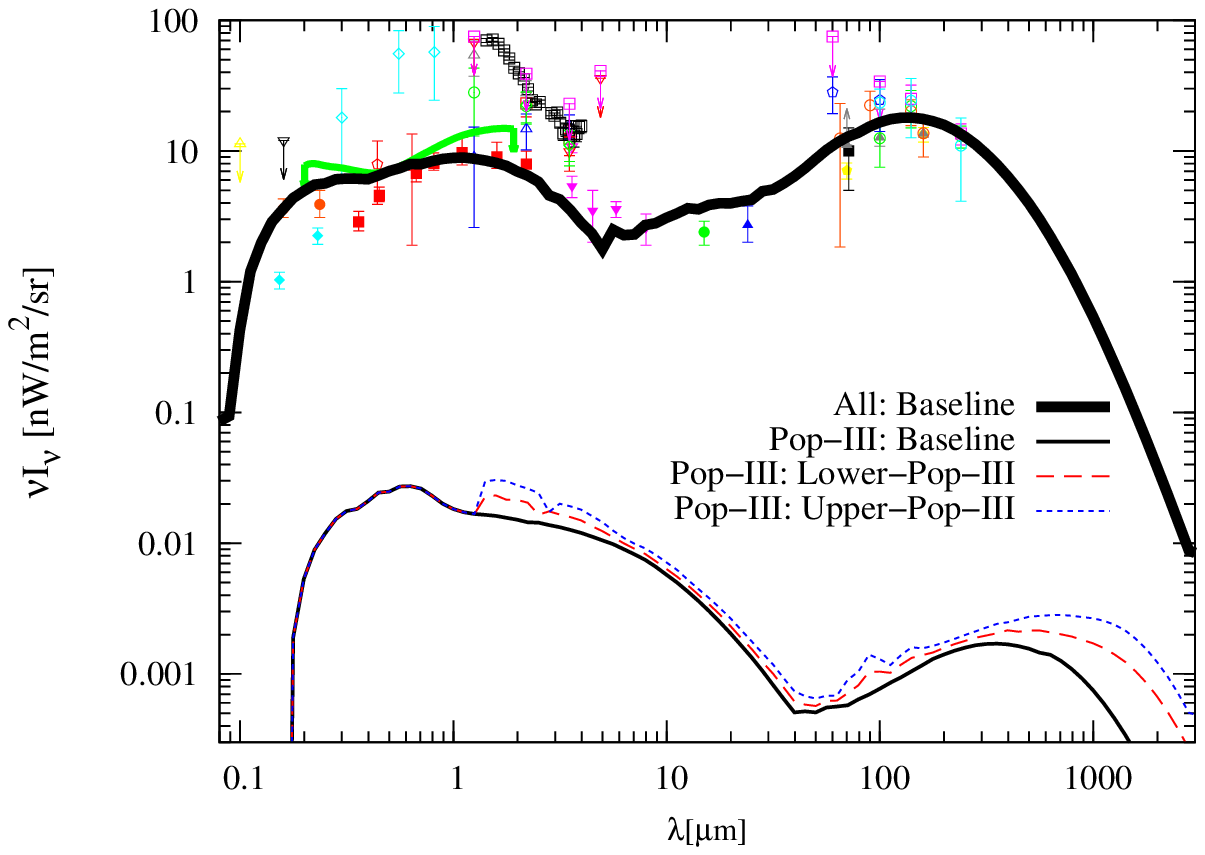} 
\caption{Same as Fig. \ref{fig:ebl}, but showing the Pop-III contribution to the EBL. Thick solid curve shows the total EBL. Thin-solid, dashed, and dotted curve corresponds to the baseline, the upper-Pop-III and the lower-Pop-III model, respectively. \label{fig:ebl_popIII} }
\end{center}
\end{figure}

\begin{figure*}
\begin{center}
\plotone{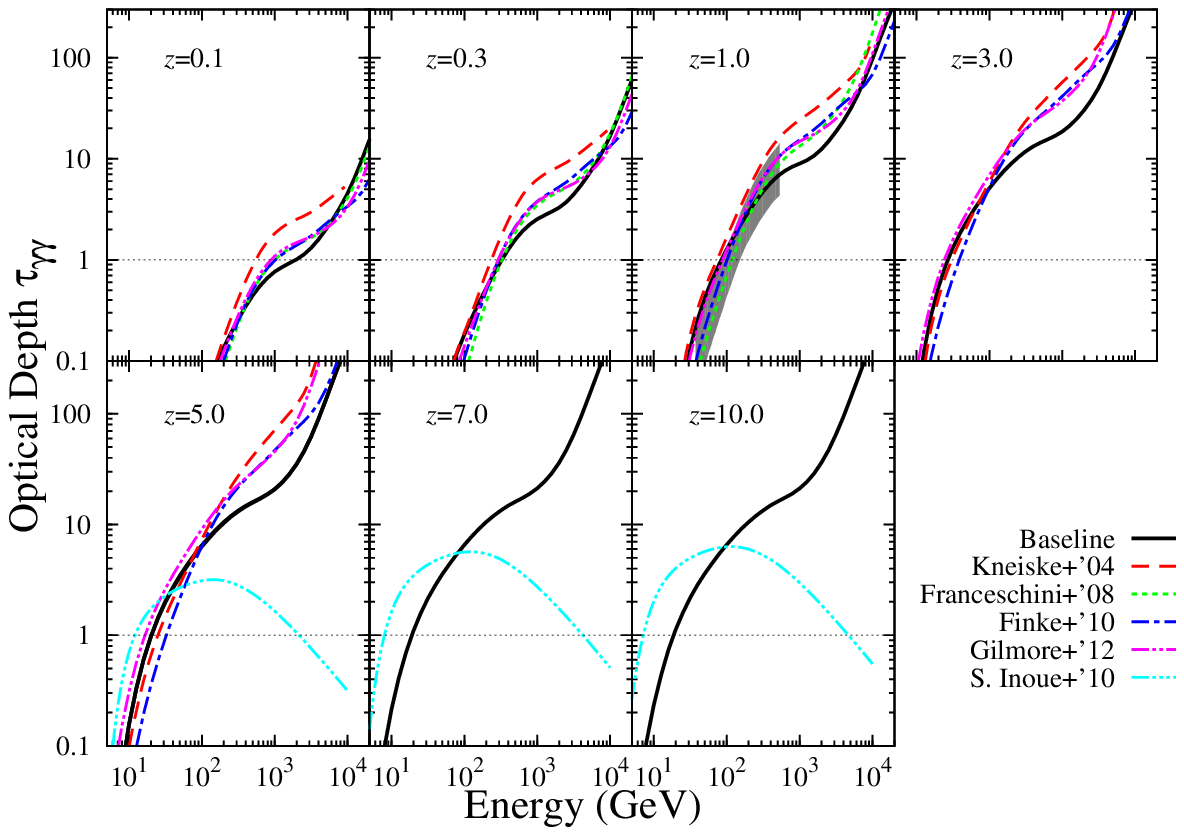} 
\caption{Optical depth to $\gamma\gamma$ interactions for observed gamma-ray energy $E_\gamma$ and sources at $z=0.1$, 0.3, 1.0, 3.0, 5.0, 7.0, and 10.0. Solid curves show our baseline model, which is nearly indistinguishable from our models including Pop-III stars. Dashed, dotted, dot-dashed, double dot-dashed, triple dot-dashed curves show the models by \citet{kne04} \citet{fra08}, \citet{fin10}, \citet{gil12}, and \citet{sin10}, respectively. The shaded region represents the 95\% confidence level measurement of the gamma-ray opacity by {\it Fermi} at $z\approx1$ \citep{abd12}. The horizontal thin dotted line marks $\tau_{\gamma\gamma}=1$. \label{fig:tau_gamma} }
\end{center}
\end{figure*}

The overall shape of our EBL model is consistent with the observational data. Our model is in good agreement with the observations by {\it Pioneer 10/11} \citep[][open pentagon symbols in Fig. \ref{fig:ebl}]{mat11_pioneer} which directly measured the EBL from outside the zodiacal region. It also does not violate the limits from gamma-ray observations \citep{aha06, alb08_3c279}.

Compared to other models, we tend to predict more photons at $\lambda\le0.4{\rm \mu m}$ and less photons at $\lambda>0.4{\rm\mu m}$. In the UV range, all other models except for \citet{ste06} are consistent with the {\it GALEX} data \citep{xu05}, while ours are consistent with the {\it HST} data \citep{gar00}. Both observational data points show the galaxy counts integrated down to zero luminosity. One reason for this difference may be in the treatment of dust obscuration. The Calzetti law \citep{cal00} that we use was shown by
\citet{som12} to result in more UV photons compared to the multi-dust component model adopted by \citet{gil12,som12}. We also note that the data points by \citet{mad00} are galaxy counts integrated down to the detection limit of the {\it HST}, implying a weaker lower-limit to the EBL.

At $0.4{\rm \mu m} < \lambda < 10{\rm \mu m}$, our model is in reasonably good agreement with \citet{kne10} who provide lower-limits to the EBL. The CSFHs in our model and in \citet{kne10} are a factor of two to three lower than that of \citet{hop06} that were used in most previous studies. We note that \citet{kne10} only discussed the global average over the Universe and did not account for the distributions of metallicity and dust attenuation in different galaxies.

Fig. \ref{fig:ebl_popIII} shows the Pop-III contribution to the EBL in our models, which are all $\le0.03$ nW  m$^{-2}$ sr$^{-1}$ and less} than 0.5\% of the total NIR background radiation. It is far too low to explain the {\it IRTS} data \citep{mat05}, even at the highest levels allowed by the reionization constraints \cite[c.f.][]{Fer12}. Moreover, even if the ionizing photons from Pop III stars that are absorbed inside galaxies are converted to Ly--$\alpha$ photons, the NIR flux will increase by only 15\% in case B recombination. Therefore, the NIR background is unlikely to provide strong constraints on Pop III stars, at least in the framework of our model.
Note that our Pop-III EBL spectrum shows two peaks,
the one in the optical caused by the minor population at low $z$,
and one in the NIR due to the redshifted, enhanced population at $z>z_c=10$.
For the upper-Pop-III model, the contribution from dust in the FIR
can be a few percent of the total FIR EBL at $\lambda\gtrsim1000 \mathrm{\mu m}$.

\section{Gamma-ray Attenuation}
\label{sec:tau}
\subsection{Gamma-ray Opacity}

\begin{figure*}
\begin{center}
\plotone{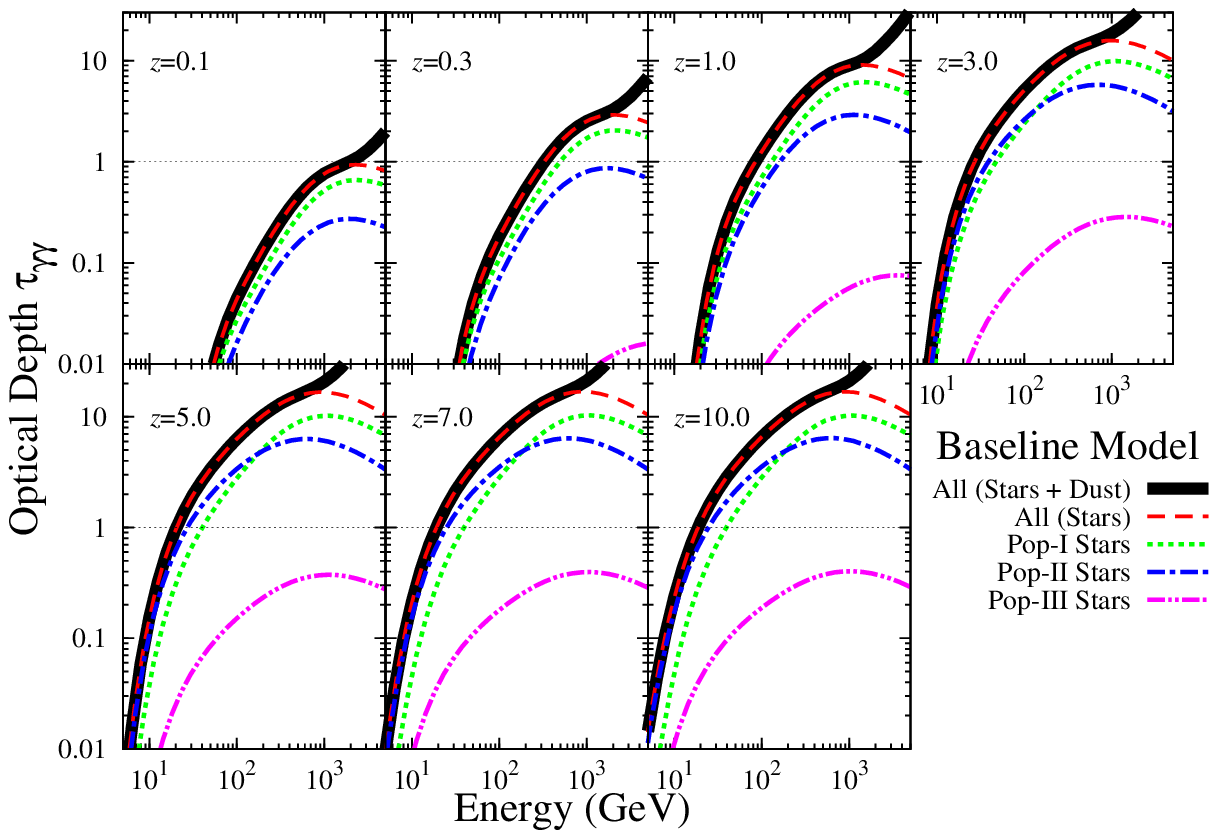} 
\caption{Same as Fig. \ref{fig:tau_gamma}, but separately for each stellar population. Solid, dashed, dotted, dot-dashed, and double dot-dashed curves show the contributions to the baseline model from all stars plus dust, all stars, Pop-I stars, Pop-II stars, and Pop-III stars, respectively. The contribution of Pop-III stars is small and does not appear in the panel for $z=0.1$.
\label{fig:tau_gamma_pop} }
\end{center}
\end{figure*}

From the redshift-dependent intensity of the EBL as given in \S\ref{sec:ebl},
we can compute the opacity for high-energy gamma rays to $\gamma\gamma$ pair production interactions. The cross section for this process is \citep{hei54}
\begin{eqnarray}
\sigma_{\gamma\gamma}(E_\gamma,\epsilon,&\theta&)=\frac{3\sigma_T}{16}(1-\beta^2) \nonumber \\
&&\times\left[2\beta(\beta^2-2)+(3-\beta^4)\ln\left(\frac{1+\beta}{1-\beta}\right)\right],
\end{eqnarray}
where $\epsilon$ is the energy of the background photon, $E_\gamma$ is the energy of the propagating high energy photon, and $\beta$ is 
\begin{equation}
\beta\equiv\sqrt{1-\frac{2m_e^2c^4}{\epsilon E_\gamma(1-\cos\theta)}};\ \ \mu\equiv\cos\theta.
\end{equation}
where $\theta$ is the angle between the two colliding photons. 
The photon energy for which the cross section peaks is given by Eq. \ref{eq:ene_ebl}.

For a photon emitted by a source at redshift $z_s$ and observed at $z=0$ with energy $E_\gamma$, the contribution to the $\gamma\gamma$ optical depth between $z_s$ and $z_0$ ($0<z_0<z_s$) is
\begin{eqnarray}
\label{eq:tau_gg}
\tau_{\gamma\gamma}(E_\gamma,z_{\rm 0}, z_s)&=&\int_{z_{\rm 0}}^{z_s}dz\int_{-1}^{1}d\mu\int_{\epsilon_{\rm th}}^{\infty}d\epsilon\frac{dl}{dz}\frac{1-\mu}{2} \nonumber \\ 
&&\times \frac{dn(\epsilon,z)}{d\epsilon}\sigma_{\gamma\gamma}(E_\gamma(1+z),\epsilon,\theta),
\end{eqnarray}
where $\epsilon_{\rm th}$ is the pair production threshold energy,
\begin{equation}
\epsilon_{\rm th}=\frac{2m_e^2c^4}{E_\gamma(1+z)(1-\mu)}.
\end{equation}

\begin{figure*}
\begin{center}
\plotone{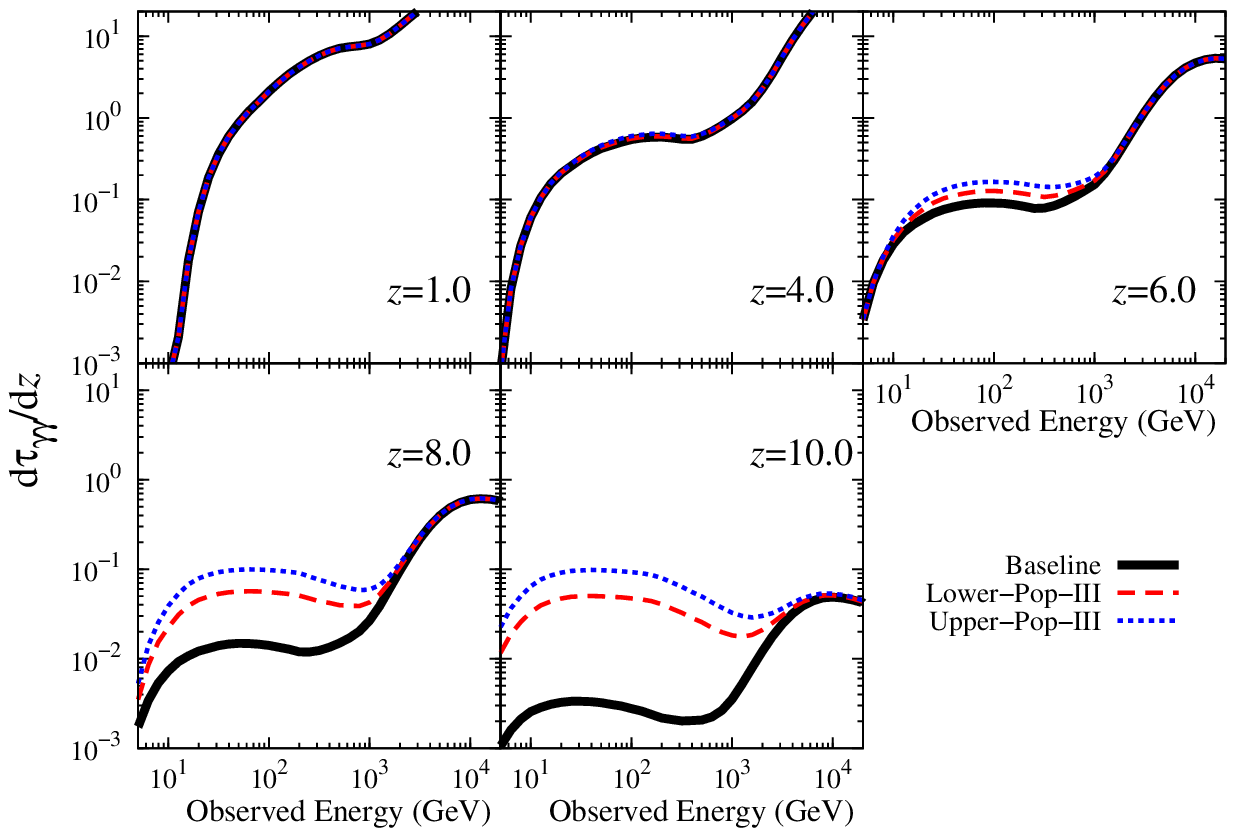} 
\caption{Differential $\gamma\gamma$ optical depth $d\tau_{\gamma\gamma}/dz$ with respect to redshift $z$ for observed gamma-ray energy $E_\gamma$ and sources at $z=1.0$, 4.0, 6.0, 8.0, and 10.0. Solid, dashed, and dotted curves show the baseline, lower-Pop-III, and upper-Pop-III models, respectively. \label{fig:dtau_gamma} }
\end{center}
\end{figure*}

Fig. \ref{fig:tau_gamma} shows the $\gamma\gamma$ optical depth as a function of the observed gamma-ray energy $E_\gamma$ for sources at selected redshifts $z=0-10$, compared with various previous models \citep{kne04,fra08,fin10,gil12,sin10}.
For all models, $z_0=0$ in Eq. \ref{eq:tau_gg}, except for \citet{sin10} where $z_0=4$, the minimum redshift in this model. For our model, only the baseline case is shown, since the Pop-III contribution turns out to be nearly indistinguishable (see below and \S. \ref{sec:ebl}). The detailed output for the $\gamma\gamma$ optical depth are publicly available at our website\footnote{\url{http://www.slac.stanford.edu/\%7eyinoue/Download.html}}. Absorption by the CMB photons is not included here. As described in \S. 2.2, there are uncertainties in the redshift evolution of our dust emissivity model in the MIR--FIR, and consequently also in $\tau_{\gamma\gamma}$ above several TeV at $z=0$. However, at these energies, the opacity due to stellar emission is already of order $\sim10$ (See Fig. \ref{fig:tau_gamma_pop}) and will likely mask such uncertainties.

Our model is consistent with the measurements of gamma-ray opacity by {\it Fermi} at $z\approx1$ at the 95\% confidence level \citep{abd12}. Although it is also generally consistent with previous models \citep{kne04,fra08,fin10,gil12} at $E_{\gamma}\lesssim400/(1+z)$ GeV for $z\le5$, the opacity above $E_{\gamma}\sim400/(1+z)$ GeV is a factor of $\sim2$ lower. We recall that our local EBL is lower at $\lambda>0.4 \ {\rm {\mu m}}$ (Fig. \ref{fig:ebl}), corresponding to $\gamma\gamma$ interactions preferentially with $\gtrsim300$ GeV photons (Eq. \ref{eq:ene_ebl}).

For very high-redshift sources at $z \gtrsim 6$, we expect spectral attenuation above $\sim$20 GeV.
This is appreciably higher than in the model of \citet{sin10}, who suggested $\sim12$ GeV at $z\sim5$ and $\sim6-8$ GeV at $z\gtrsim8-10$. Their basis was the models of cosmic reionization by \citet{cho06, cho09}, which included Pop III stars as well as QSOs and were developed to explain essentially all observational constraints related to reionization, including $\tau_e$ from {\it WMAP} and $x_{\rm HI}$ from GP measurements. However, being optimized for the reionization epoch, they focused on $z \ge 4$ and did not account for Population-I stars with $Z>0.02 Z_\odot$ nor dust. While a thorough comparison between the two models is not feasible,
the principal difference appears to be in the CSFH for Pop-II stars, which is a factor of $\sim 3-10$ higher at $z \gtrsim 6$ in \citet{sin10} compared to our baseline model here. This demonstrates that such differences in the CSFHs can be clearly distinguishable through future gamma-ray observations.

The highest redshifts of high-energy gamma-ray sources known so far are $z \sim 3$ for blazars and $z=4.35$ for GRBs. Based on a model for the gamma-ray luminosity function of blazars, \citet{ino10_highz} proposed that {\it Fermi} may eventually detect blazars up to $z\sim6$. GRBs are known to occur at $z > 6$ \citep{kaw06,gre09}, at least up to $z \sim8.2$ \citep{tan09,sal09}, and probably out to the epoch of first star formation in the Universe \citep{bro06}. If a bright burst similar to GRB 080916C \citep{abd09_080916C} occurs at such redshifts, CTA may be able to measure its spectrum up to $z \sim 7-10$ \citep{sin12_cta} (and possibly even higher, see e.g. \citet{tom11}),
offering a unique probe of the EBL during cosmic reionization.

Fig. \ref{fig:tau_gamma_pop} shows the $\gamma\gamma$ opacity due to each stellar population separately for the baseline model: Pop-I stars ($10^{-2.5}\le Z$), Pop-II stars ($10^{-4}\le Z<10^{-2.5}$), and Pop-III stars ($Z<10^{-4}$). Here we have chosen the dividing metallicity between Pop-II halo stars and Pop-I disk stars to be $Z=10^{-2.5}$, as the distinction between the two populations is known to occur at [Fe/H]$\simeq-1$, corresponding to $Z\simeq10^{-2.7}-10^{-2.3}$ \citep{whe89,mcw97,pro00}.
Although the gamma-ray attenuation signature of Pop-III stars seems difficult to discern, that due to Population II stars should be observable in future observations of high-$z$ gamma-ray sources and will provide a valuable probe of the evolving UV EBL in the cosmic reionization epoch. The detection of even one photon from such redshifts
will impose useful limits on cosmic reionization models as well as Pop III stars.

Fig. \ref{fig:dtau_gamma} shows the differential contributions to the $\gamma\gamma$ optical depth $d\tau_{\gamma\gamma}/dz$ with respect to redshift in our models. Since the Pop-III component is enhanced only at $z>10$, its effect at $z\lesssim6$ is insignificant, while at $z\gtrsim6$, differences can been seen of $\sim$3\%, $\sim$10\% and $\sim$20\%  at 20 GeV relative to the baseline model at $z=6$, 8, and 10, respectively. Discrimination between the models would be possible only if differences of $\sim10$\% in flux can be identified.

\subsection{Comparison with Current GeV \& TeV data}

Gamma-ray astronomy has seen enormous progress during the last decade, led by new generation facilities such as {\it Fermi}, H.E.S.S., MAGIC, and VERITAS, among others. Further progress is anticipated in the near future with CTA. CTA is expected to detect $>100$ blazars up to $z\sim2.5$ \citep{ino10a,ino11_cta,sol12}. It is also expected to detect GRBs at a rate of order a few per year, possibly out to much higher redshifts \citep{kak12,gil12_cta,sin12_cta}. Such observations will allow us to greatly clarify the evolution of the EBL in the UV-NIR bands. Detailed observations of TeV blazars at low $z$ will also be crucial for probing the FIR EBL, which would not be possible with high-$z$ sources. Starburst galaxies have also been suggested as alternative targets for studying the FIR EBL \citep{dwe12}, even though internal gamma-ray absorption may limit their usefulness \citep[e.g.][]{ino11_gal}.

\begin{figure}
\begin{center}
\plotone{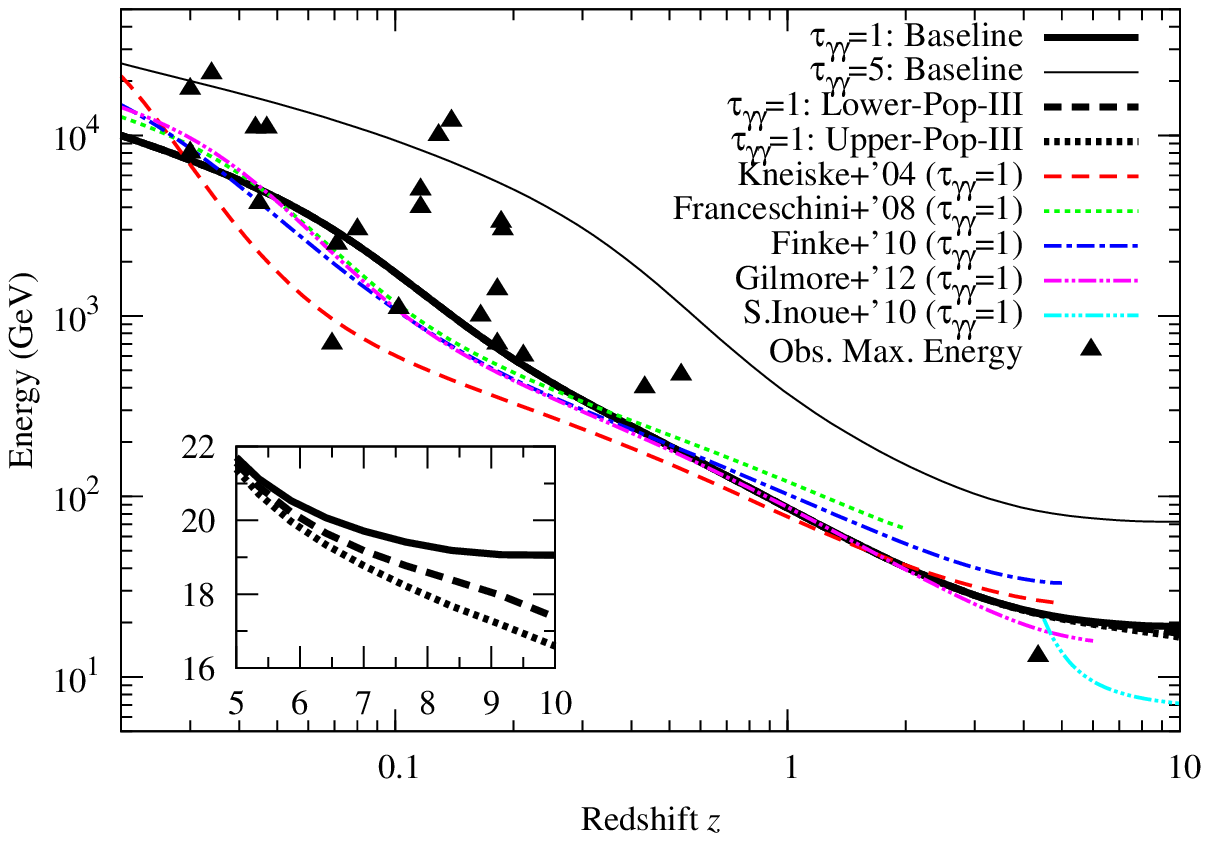} 
\caption{Gamma-ray horizon energy where $\tau_{\gamma\gamma}=1$. The baseline, lower-Pop-III, and upper-Pop-III models are shown by the thick solid, thick dashed, and thick dotted curves, respectively. Dashed, dotted, dot-dashed, double dot-dashed, triple dot-dashed curves represent the models by \citet{kne04} \citet{fra08}, \citet{fin10}, \citet{gil12}, and \citet{sin10}, respectively. The thin solid curve shows the case of $\tau_{\gamma\gamma}=5$ for the baseline model. The filled data points are the observed maximum energies of photons from a sample of blazars \citep{fin09} and GRB 080916 C \citep{abd09_080916C}. Since other papers do not cover the opacity at $z=0-10$, we do not show the opacity of each model at the outside of the redshift range of each paper. The small panel in the figure shows the gamma-ray opacity horizon at $z=5-10$.
\label{fig:tau_horizon} }
\end{center}
\end{figure}

The gamma-ray horizon energy at which $\tau_{\gamma\gamma}=1$ as a function of $z$, known as the Fazio--Stecker relation \citep{faz70}, is shown in Fig. \ref{fig:tau_horizon} in comparison with other models \citep{kne04,fra08,fin10,gil12,sin10}. We also plot the maximum energies of photons detected from a sample of blazars \citep[see][for a list and references]{fin09} as well as GRB 080916C \citep{abd09_080916C}. The inset in Fig. \ref{fig:tau_horizon} is a blow up for $z=5-10$ to emphasize the differences among our models with and without Pop-III stars.

The highest photon energies for many blazars lie in regions considerably above the $\tau_{\gamma\gamma}=1$ curves for all EBL models, indicating that their spectra are likely to be highly attenuated. 
It is also clear that GRB 080916C provides an important constraint on the EBL at $z \sim 4$. Our model predicts that the Universe is transparent below 20 GeV even at $z>4$. 

\begin{figure*}
\begin{center}
\plotone{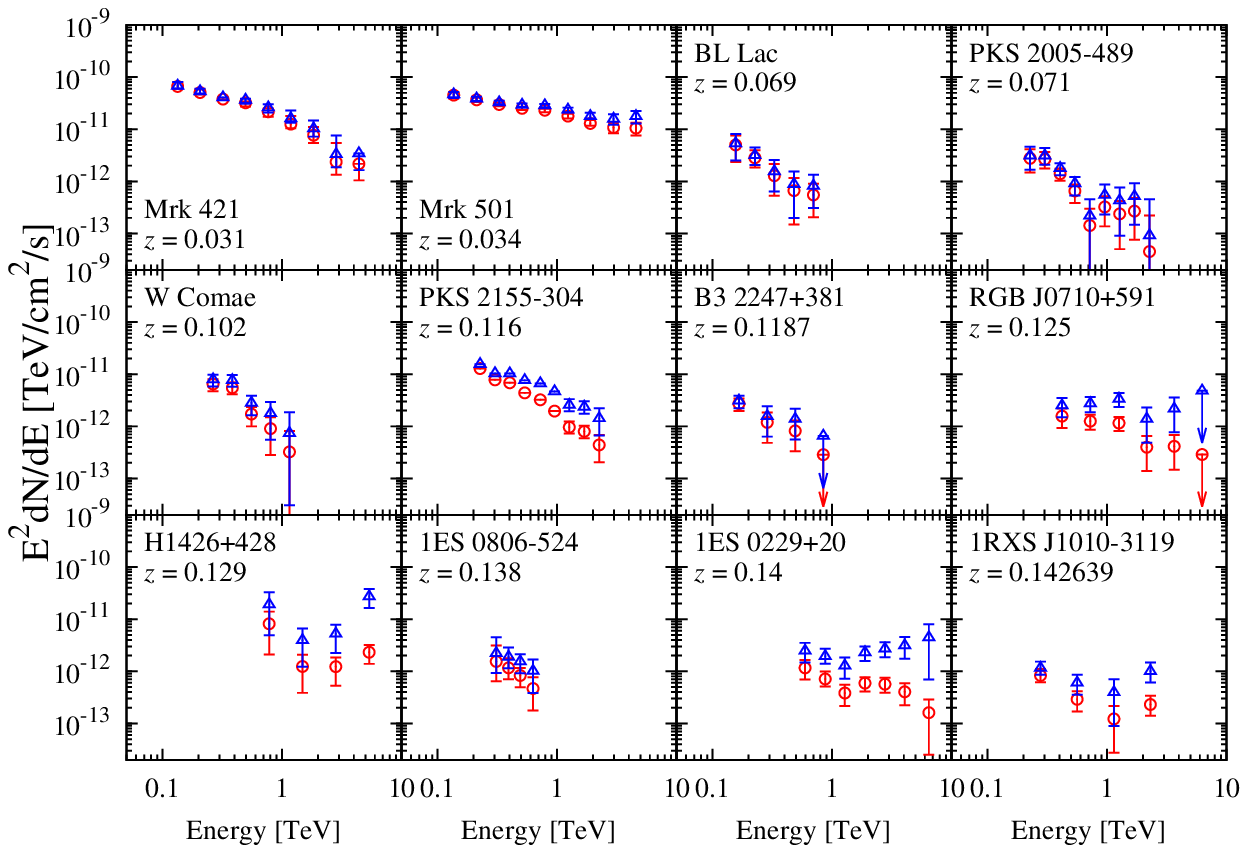} 
\caption{Spectra of TeV blazars at $z\le0.15$ as observed (circle) and inferred before EBL attenuation (triangle) with our baseline model. References for the data are 
Mrk 421 \citep{alb07_mrk421}
, Mrk 501 \citep{alb07_Mrk501}
, BL Lac \citep{alb07_BLLac}
, PKS 2005-489 \citep{aha05_PKS2005-489}
, W Comae \citep{acc08_WComae}
, PKS 2155-304 \citep{aha05_PKS2155-304}
, B3 2247+381 \citep{ale12_B32247+381}
, RGB J0710+591 \citep{acc10_RGBJ0710+591}
, H 1426+428 \citep{aha02_H1426+428}
, 1ES 0806-524 \citep{acc09_1ES0806+524}
, 1ES 0229+200 \citep{aha07_1ES0229+200}
, 1RXS J1010-3119\citep{abr12_1RXSJ101015.9-311909}.
  \label{fig:tau_zb1.5} }
\end{center}
\end{figure*}

\begin{figure*}
\begin{center}
\plotone{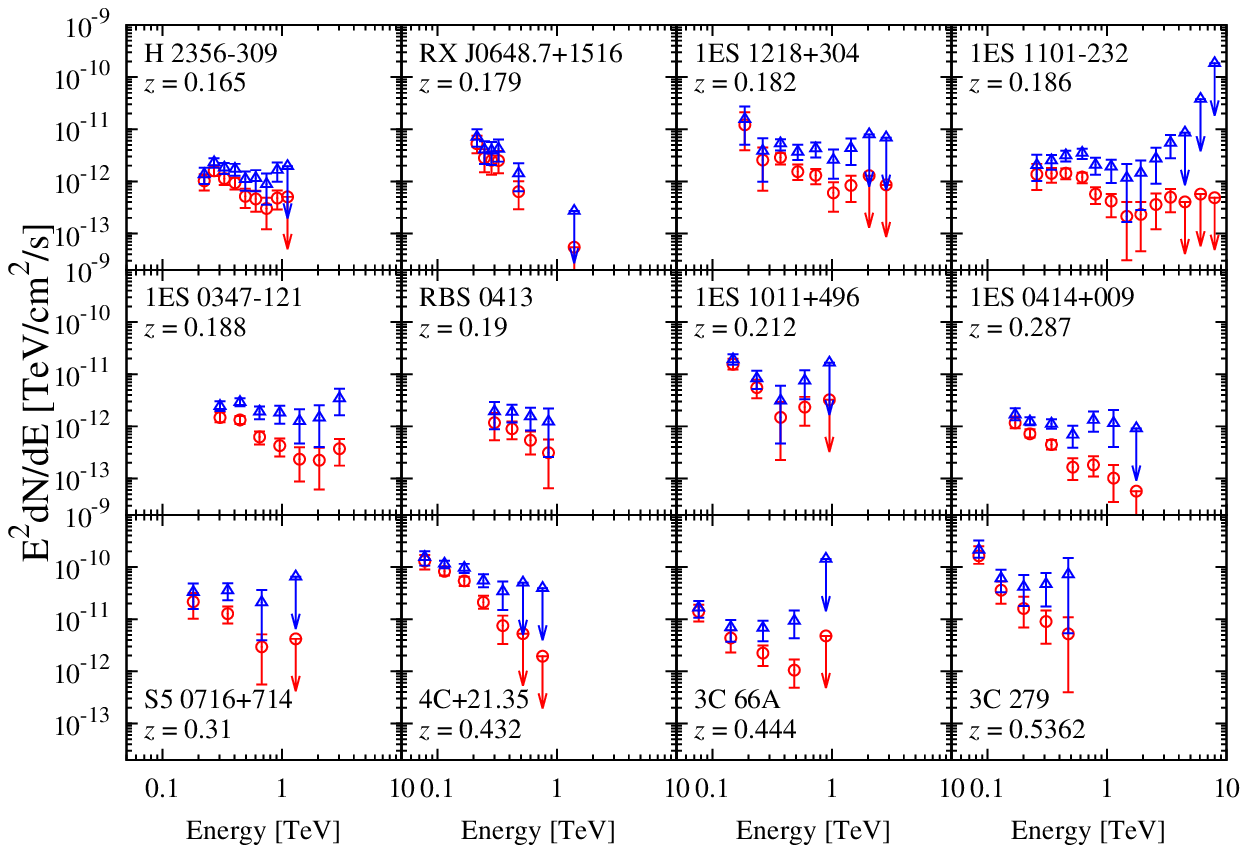} 
\caption{Same as Fig. \ref{fig:tau_zb1.5}, but for sources at $z>0.15$. References for the data are 
H 2356-309 \citep{aha06_H2356-309}
, RX J0648.7+1516 \citep{ali11_RXJ0648.7+1516}
, 1ES 1218+304 \citep{acc09_1ES1218+304}
, 1ES 1101-232 \citep{aha07_1ES1101-232}
, 1ES 0347-121 \citep{aha07_1ES0347-121}
, RBS 0413 \citep{ali12_RBS0413}
, 1ES 1011+496 \citep{alb07_1ES1011+496}
, 1ES 0414+009 \citep{abr12_1ES0414+009}
, S5 0716+714 \citep{and08_S50716+714}
, 4C+21.35 \citep{ale11_PKS1222+21}
, 3C 66A \citep{ale11_3C66A}
, 3C 279 \citep{alb08_3c279}.
  \label{fig:tau_za1.5} }
\end{center}
\end{figure*}

Figs. \ref{fig:tau_zb1.5} and \ref{fig:tau_za1.5} show the observed spectra of TeV blazars at $z\le0.15$ and $z>0.15$, respectively, together with their intrinsic spectra before attenuation by the EBL, assuming our baseline model. If the TeV emission from these sources originate from electrons accelerated according to the simplest, test-particle theory of diffusive shock acceleration \citep{bla87}, the hardest spectrum is expected to be of photon index $\Gamma_{\rm int}=1.5$. Although the intrinsic spectra inferred here are generally consistent with $\Gamma_{\rm int} \ge 1.5$, some sources such as 1ES 0229+20 and 1ES 1101-232 show evidence of harder spectra above several hundred GeV \citep[see also][]{fin10}.

To explain such intrinsically hard spectra, some authors have recently suggested secondary cascade components  generated by very high energy cosmic-rays or gamma-rays, which may also offer a probe of intergalactic magnetic fields \citep[e.g.][]{ess10,ess11,ess12,mur12,aha12}. Others have proposed effects of time-dependence, stochastic acceleration or multiple emission components \citep{lef11,lef11_blob}. Future CTA observations of these objects with high energy and time resolution will elucidate such issues.

The signature of EBL absorption has not been seen in the spectrum of the extragalactic gamma-ray background (EGB) above 100 GeV \citep{ack11_TeVPA}, even though it is naturally expected if its origin is cosmological \citep{ino11,ino12}. By considering the effects of cascade emission, \citet{ino12} have recently shown that 
if the EGB at <100 GeV \citep{abd10_egrb} is entirely composed of known types of sources whose spectra are well constrained by existing observations, the measured EGB at >100 GeV would be inconsistent with this hypothesis, even for a low EBL such as proposed here. Further detailed spectral studies of extragalactic gamma-ray sources are required to resolve this issue.

\section{Conclusions}
\label{sec:con}

We have developed models for the EBL over the redshift range $z=10$ to $z=0$ on the basis of a semi-analytical model of hierarchical galaxy formation, into which Pop-III stars were incorporated in a simplified fashion. Our baseline model is consistent with a wide variety of observational data for galaxies below $z\sim6$ \citep{nag04,kob07,kob10}, and is also capable of reionizing the Universe by $z<8$. However, in order to account for the Thomson scattering optical depth measured by {\it WMAP}, the ionizing photon emissivity is required to be 50-100 times higher at $z>10$. This is line with recent observations of galaxy candidates at $z\sim8$, as long as the contribution from faint galaxies below the sensitivity of current telescopes is not large \citep[e.g.][]{bou12}. The ``missing'' ionizing photons may possibly be supplied by Pop-III stars forming predominantly at these epochs in sufficiently small galaxies.

The EBL intensity at $z=0$ in our model is generally not far above the lower limits derived from galaxy counts. Our model is also in good agreement with the data from {\it Pioneer} \citep{mat11_pioneer} directly measured from outside the zodiacal region. The Pop-III contribution to the NIR EBL is $\le0.03$ nW  m$^{-2}$ sr$^{-1}$, less than 0.5 \% of the total in this band, even at the maximum level compatible with {\it WMAP} measurements. The putative NIR EBL excess \citep{mat05}, which also conflicts with the upper limits from gamma-ray observations \citep{aha06}, may have a zodiacal origin rather than Pop-III stars.

Up to $z\sim3$--5, the $\gamma\gamma$ opacity in our model is comparable to that in the majority of previously published models \citep{kne04,fra08,fin10,gil12} below $E_\gamma\sim400/(1+z)$ GeV, while it is a factor of $\sim2$ lower above this energy. The Universe is predicted to be largely transparent below 20 GeV even at $z>4$.

Estimates based on the observed gamma-ray luminosity function of blazars show that {\it Fermi} may detect blazars up to $z\sim6$ \citep{ino10_highz}. CTA may possibly detect GRBs up to similar redshifts \citep{sin12_cta}. However, the contribution of Pop-III stars 
may be difficult to discern in the attenuated spectra of high-redshift gamma-ray sources, even at the highest levels allowed by the WMAP constraints. Nevertheless, the signature of Population II stars is expected to be observable in high-$z$ gamma-ray sources, providing a unique and valuable probe of the evolving EBL in the rest-frame UV.

\acknowledgements
We thank the anonymous referee for useful comments and suggestions. We also thank Floyd Stecker and Alberto Dominguez for helpful comments and Tirth Roy Choudhury for providing numerical data from his model. YI acknowledges support by the Research Fellowship of the Japan Society for the Promotion of Science (JSPS). 
SI is supported by Grants-in-Aid Nos. 22540278 and 24340048 from MEXT of Japan.

\appendix
\section{Cosmic Star Formation History}
\label{sec:app1}
In this appendix, we briefly review the method of conversion to the CSFH from the measurement of the 1500 {\AA} UV LD $\varrho_{\mathrm{obs}}(z)$, for which there are two steps \citep[see][for details]{kob12}. One is the correction for dust obscuration to derive the intrinsic LD $\varrho_{\mathrm{int}}(z)$ and the other is the conversion of the intrinsic LD to the star formation rate density $\dot\rho_{\mathrm{star}}(z)$. Dust obscuration correction and star formation rate conversion are given by
\begin{eqnarray}
\varrho_{\mathrm{int}}(z) &=& C_\mathrm{dust}(z)\varrho_\mathrm{obs}(z),\\
\dot\rho_\mathrm{star}(z) &=& C_\mathrm{SFR}(z)\varrho_\mathrm{int}(z).
\end{eqnarray}

\citet{hop04} assumed redshift-independent $C_\mathrm{dust}$ and  $C_\mathrm{SFR}$, while \citet{bou07} and \citet{cuc12} assumed redshift-dependent $C_\mathrm{dust}$ but with  redshift-independent $C_\mathrm{SFR}$ \citep[see][for details]{bou07,cuc12}. However, these simple assumptions for all redshifts can cause an overestimation of the CSFH. \citet{kob12} have recently proposed a new redshift dependent conversion method based on their semi-analytical galaxy formation model as
\begin{eqnarray}
C_\mathrm{dust}(z) &=& 2.983 \exp[-0.3056(1+z)]+1,\\
C_\mathrm{SFR}(z)&=& 10^{-28.01}[1-5.915\times10^{-5}(1+z)+7.294\times10^{-4}(1+z)^2] \ \mathrm{M_\odot \ yr^{-1}\ (erg\ s^{-1}\ Hz^{-1})^{-1}}.
\end{eqnarray}

\begin{figure}
\begin{center}
\plotone{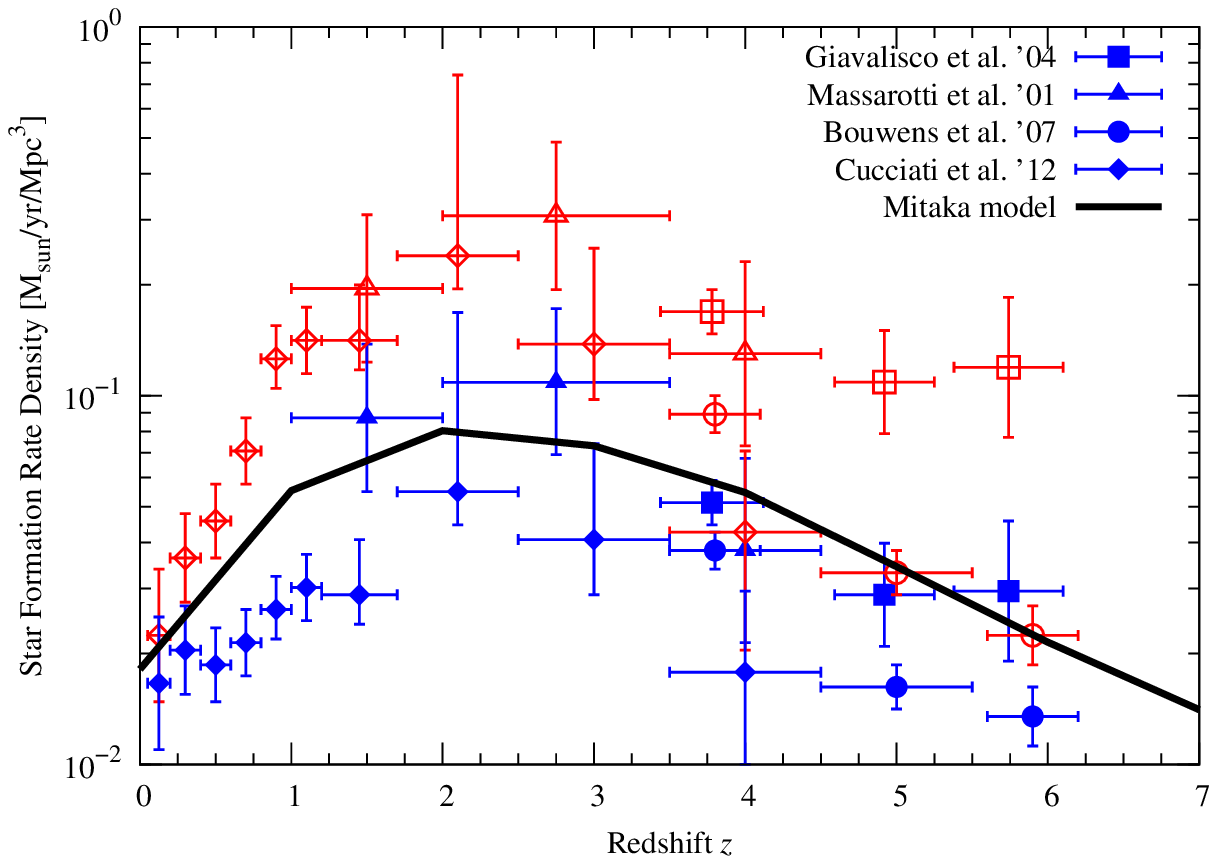} 
\caption{Cosmic star formation history. The solid curve shows the total in the baseline Mitaka model. Square, triangle, circle, and diamond symbols show the expected CSFH from the observed 1500 {\AA} LD by \citet{gia04}, \citet{mas01}, \citet{bou07},  and \citet{cuc12} respectively. Open symbols are based on the conversion method by \citet{hop04} for square and triangle symbols, by \citet{bou07} for circle symbols, and by \citet{cuc12} for diamond symbols. Filled symbols are based on the method by \citet{kob12}. We converted the cosmology assumed in the original references to that assumed here.
  \label{fig:CSFRD_K13} }
\end{center}
\end{figure}

Fig. \ref{fig:CSFRD_K13} shows the CSFH converted from the observed 1500 {\AA} LD using the methods developed by \citet{hop04}, \citet{bou07}, \citet{cuc12}, and \citet{kob12}. The 1500 {\AA} UV LD data used in \citet{hop04} are from \citet{gia04} and \citet{mas01}. The expected CSFH from our baseline Mitaka model fits well to the CSFH following \citet{kob12}.

\end{document}